\documentclass[10pt]{article}

\usepackage{bbm,bm}
\usepackage{amsmath,amsfonts,amsthm,amssymb}
\usepackage{graphicx,float,wrapfig}
\usepackage[top=1in, bottom=1in, left=1in, right=1in]{geometry}

\emergencystretch=\maxdimen
\usepackage{color}

\newtheorem{thm}{Theorem}[section]
\newtheorem{cor}[thm]{Corollary}
\newtheorem{lem}[thm]{Lemma}

\newtheorem{defn}[thm]{Definition}
\theoremstyle{remark}

\newcommand{\e}{\begin{equation}}
\newcommand{\ee}{\end{equation}}

\title{Approximating Sampled Sinusoids and Multiband Signals Using Multiband Modulated DPSS Dictionaries}
\author{Zhihui Zhu and Michael B. Wakin\thanks{Email: zzhu,mwakin@mines.edu. This work was supported by NSF grant CCF-1409261.}\\
  \small Department of Electrical Engineering and Computer Science\\ \small Colorado School of Mines}

\date{}

\begin{document}
\maketitle

\abstract
Many signal processing problems---such as analysis, compression, denoising, and reconstruction---can be facilitated by expressing the signal as a linear combination of atoms from a well-chosen dictionary. In this paper, we study possible dictionaries for representing the discrete vector one obtains when collecting a finite set of uniform samples from a multiband analog signal. By analyzing the spectrum of combined time- and multiband-limiting operations in the discrete-time domain, we conclude that the information level of the sampled multiband vectors is essentially equal to the time-frequency area. For representing these vectors, we consider a dictionary formed by concatenating a collection of modulated Discrete Prolate Spheroidal Sequences (DPSS's). We study the angle between the subspaces spanned by this dictionary and an optimal dictionary, and we conclude that the multiband modulated DPSS dictionary---which is simple to construct and more flexible than the optimal dictionary in practical applications---is nearly optimal for representing multiband sample vectors. We also show that the multiband modulated DPSS dictionary not only provides a very high degree of approximation accuracy in an MSE sense for multiband sample vectors (using a number of atoms comparable to the information level), but also that it can provide high-quality approximations of all sampled sinusoids within the bands of interest.

~


~


\section{Introduction}

\subsection{Signal dictionaries and representations}

Effective techniques for signal processing often rely on meaningful representations that capture the structure inherent in the signals of interest. Many signal processing tasks---such as signal denoising, recognition, and compression---benefit from having a {\em concise} signal representation. Concise signal representations are often obtained by ($i$) constructing a {\em dictionary} of elements drawn from the signal space, and then ($ii$) expressing the signal of interest as a linear combination of a small number of atoms drawn from the dictionary.

Throughout this paper, we consider the signal space $\mathbb{C}^{N}$, and we represent a dictionary as an $N\times L$ matrix $\bm{\Psi}$, which has columns (or atoms) $\bm{\psi}_0, \bm{\psi}_1, \ldots, \bm{\psi}_{L-1}$. Using this dictionary, a signal $\bm{x}\in\mathbb{C}^{N}$ can be represented exactly or approximately as a linear combination of the $\bm{\psi}_i$:
\e
\bm{x} \approx \bm{\Psi}\bm\alpha = \sum_{i=0}^{L-1}\bm\alpha[i]\bm{\psi}_i
\nonumber\ee
for some $\bm{\alpha}\in\mathbb{C}^L$, whose entries are referred to as coefficients.

When the coefficients have a small fraction of nonzero values or decay quickly, one can form highly accurate and concise approximations of the original signal using just a small number of atoms. In some cases, one can achieve this using a linear approximation that is formed with a prescribed subset of $J<L$ atoms:
\e
\bm{x}\approx \sum_{i\in\Omega}\bm \alpha[i]\bm{\psi}_i,
\label{eq:linearApprox}\ee
where $\Omega\subset\{0,1,\ldots,L-1\}$ is a fixed subset of cardinality $J$. For example, one might use the lowest $J$ frequencies to approximate bandlimited signals in a Fourier basis.

In other cases, it may be beneficial to adaptively choose a set of atoms in order to optimally represent each signal. Such a nonlinear approximation can be expressed as
\e
\bm{x}\approx \sum_{i\in\Omega(\bm{x})}\bm \alpha[i]\bm{\psi}_i,
\label{eq:nonlinearApprox}\nonumber\ee
where $\Omega(\bm{x})\subset\{0,1,\ldots,L-1\}$ is a particular subset of cardinality $J$ and can change from signal to signal. A more thorough discussion of this topic, which is also known as {\em sparse approximation}, can be found in~\cite{davis1994adaptive,Devore1998Nolinear,Mallat2008WaveletTour}. Sparse approximations have been widely used for signal denoising~\cite{donoho1995Denoising}, signal recovery~\cite{bruckstein2009sparseSolutions} and compressive sensing (CS)~\cite{CandesRombergTao2006RobustUncertaintyPrinciples,CandesWakin2008IntroductionCS,candes2006stableRecovery,cohen2009compressedSensing,donoho2006compressedSensing}, an emerging research area that aims to break through the Shannon-Nyquist limit for sampling analog signals. A challenge in finding the best $J$-term approximation for a given signal $\bm{x}$ is to identify which of the ${L \choose J}$ subspaces (or, equivalently, index sets $\Omega(\bm{x})$) to use. This problem has garnered much attention in the applied mathematics and signal processing communities, and conditions can be established under which methods based on convex optimization~\cite{CandesRombergTao2006RobustUncertaintyPrinciples,chen1998atomicDecomposition,donoho2003optimallySparseRepl1} and greedy algorithms~\cite{blumensath2008iterativeThresholding,mallat1993matchingPursuit,needell2009cosamp,tropp2004greed} provide suitable approximations.

\subsection{Dictionaries for finite-length vectors of sampled analog signals}

In this paper, we study dictionaries for representing the discrete vector one obtains when collecting a finite set of uniform samples from a certain type of analog signal. We let $x(t)$ denote a complex-valued analog (continuous-time) signal, and for some finite number of samples $N$ and some sampling period $T_s > 0$, we let
\begin{equation}
\bm{x} = \left[x(0) ~x(T_s) ~\cdots ~x((N-1)T_s)\right]^T
\label{eq:x}
\end{equation}
denote the length-$N$ vector obtained by uniformly sampling $x(t)$ over the time interval $[0,NT_s)$ with sampling period $T_s$. Here $T$ stands for the transpose operator. Our focus is on obtaining a dictionary $\bm{\Psi}$ that provides highly accurate approximations of $\bm{x}$ using as few atoms as possible.

It is the structure we assume in the analog signal $x(t)$ that motivates the search for a concise representation of $\bm{x}$. Specifically, we assume that $x(t)$ obeys a multiband signal model, in which the signal's continuous-time Fourier transform (CTFT) is supported on a small number of narrow bands (we assume the bands are known). We describe this model more fully in Section~\ref{sec:mbmodel}. Before doing so, we begin in Section~\ref{sec:mtmodel} with a simpler analog signal model for which an efficient dictionary $\bm{\Psi}$ is easier to describe.

\subsubsection{Multitone signals}
\label{sec:mtmodel}

A {\em multitone} analog signal is one that can be expressed as a sum of $J$ complex exponentials of various frequencies:
\e
x(t) = \sum_{i = 0}^{J-1} \beta_i e^{j2\pi F_i t}. \nonumber
\ee
Suppose such a multitone signal $x(t)$ is bandlimited with bandlimit $\frac{B_{\text{nyq}}}{2}$ Hz, i.e., that $\max_i|F_i|\leq \frac{B_{\text{nyq}}}{2}$. Let $\bm{x}$, as defined in \eqref{eq:x}, denote the length-$N$ vector obtained by uniformly sampling $x(t)$ over the time interval $[0,NT_s)$ with sampling period $T_s\leq \frac{1}{B_{\text{nyq}}}$ which meets the Nyquist sampling rate. We can express these samples as
\e
\bm{x}[n] = \sum_{i = 0}^{J-1} \beta_i e^{j2\pi f_i n}, ~ n = 0,1,\dots,N-1,
\label{eq:multitoneModel1}
\ee
where $f_i = F_iT_s$.
This model arises in problems such as radar signal processing with point targets~\cite{Lagunas2013JointWallMitigation} and super-resolution~\cite{candes2014towardsMathSuperResolution}.

In certain cases, an effective dictionary for representing $\bm{x}$ is the $N\times N$ discrete Fourier transform (DFT) matrix~\cite{Baraniuk2007CompressiveRadarImaging,tropp2010beyondNyquist,Lagunas2013JointWallMitigation}, where $\bm{\psi}_i[n] = e^{j2\pi in/N}$ for $i = 0,1,\dots,N-1$ and $n = 0,1,\dots,N-1$. Using this dictionary, we can write $\bm{x} = \bm{\Psi} \bm{\alpha}$, where $\bm{\alpha} \in \mathbb{C}^N$ contains the DFT coefficients of $\bm{x}$. When the frequencies $f_i$ appearing in \eqref{eq:multitoneModel1} are all integer multiples of $1/N$, then $\bm{\alpha}$ will be $J$-sparse (meaning that it has at most $J$ nonzero entries), and the sparse structure of $x(t)$ in the analog domain will directly translate into a concise representation for $\bm{x}$ in $\mathbb{C}^N$. This ``on grid'' multitone signal is sometimes assumed for simplicity in the CS literature~\cite{tropp2010beyondNyquist}. However, when the frequencies comprising $x(t)$ are arbitrary, the sparse structure in $\bm{\alpha}$ will be destroyed due to the ``DFT leakage'' phenomenon. Such a problem can be mitigated by applying a windowing function in the sampling system, as in~\cite{tropp2010beyondNyquist}, or iteratively using a refined dictionary~\cite{fannjiang2012coherence}. An alternative is to consider the model (\ref{eq:multitoneModel1}) directly as in~\cite{eftekhariRombergWakin2013matchedFiltering,tang2013compressedOffGrid}. However, such approaches cannot be generalized to scenarios in which the analog signals contain several bands, each with non-negligible bandwidth.

\subsubsection{Multiband signals}
\label{sec:mbmodel}

A more realistic model for a structured analog signal is a {\em multiband} model, in which $x(t)$ has a CTFT supported on a union of several narrow bands
\e
\mathbb{F} = \mathop{\bigcup}\limits_{i=0}\limits^{J-1}[F_i-B_{\text{band}_i}/2,F_i+B_{\text{band}_i}/2],
\nonumber\ee
i.e.,
\e
x(t) = \int_{\mathbb{F}}X(F)e^{j2\pi Ft}dF.
\nonumber\ee
Here $X(F)$ denotes CTFT of $x(t)$. The band centers are given by the frequencies $\{F_i\}_{i \in [J]}$ and the band widths are denoted by $\{B_{\text{band}_i}\}_{i \in [J]}$, where $[J]$ denotes the set $\{0,1,\ldots,J-1\}$.

Again we let $\bm{x}$, as defined in \eqref{eq:x}, denote the length-$N$ vector obtained by uniformly sampling $x(t)$ over the time interval $[0,NT_s)$ with sampling period $T_s$. We assume $T_s$ is chosen to satisfy the minimum Nyquist sampling rate, which means
\[
T_s \leq \frac{1}{B_{\textup{nyq}}}: = \frac{1}{2\max_{i \in [J]} \left\{\left|F_i\pm B_{\textup{band}_i}/2\right|\right\}}.
\]
Under these assumptions, the sampled multiband signal $\bm{x}$ can be expressed as an integral of sampled pure tones (i.e., discrete-time sinusoids)
\begin{equation}
\bm{x}[n] = \int_{\mathbb{W}}\widetilde{x}(f)e^{j2\pi fn}\; df, ~ n = 0,1,\dots,N-1,
\label{eq:xsynthesis}
\end{equation}
where the digital frequency $f$ is integrated over the union of intervals
\e
\mathbb{W} := T_s \mathbb{F} = [f_0-W_0,f_0+W_0]\cup[f_1-W_1,f_1+W_1]\cup\cdots\cup[f_{J-1}-W_{J-1},f_{J-1}+W_{J-1}] \subseteq \left[-\frac{1}{2},\frac{1}{2}\right]
\label{eq:MultiBand}
\ee
with $f_i = T_sF_i$ and $W_i = T_sB_{\text{band}_i}/2$ for all $i\in[J]$. The weighting function $\widetilde{x}(f)$ appearing in \eqref{eq:xsynthesis} equals the scaled CTFT of $x(t)$,
\[
\widetilde{x}(f) = \frac{1}{T_s} X(F)|_{F = \frac{f}{T_s}}, ~ |f| \le \frac{1}{2},
\]
and corresponds to the discrete-time Fourier transform (DTFT) of the infinite sample sequence $\{\dots,x(-2T_s),x(-T_s),x(0),x(T_s),x(2T_s),\dots\}$. (However, we stress that our interest is on the finite-length sample vector $\bm{x}$ and not on this infinite sample sequence.) Such multiband signal models arise in problems such as radar signal processing with non-point targets~\cite{AhmadQianAmin2015WallCluterDPSS} and mitigation of narrowband  interference~\cite{Davenport2010WidebandCompressiveReceiver,davenportWakin2011reconstructionCancellationDPSS}.

In this paper, we focus on building a dictionary in which finite-length sample vectors arising from multiband analog signals can be well-approximated using a small number of atoms. The DFT basis is inefficient for representing these signals because the DFT frequencies comprise only a regular, finite grid rather than a continuum of frequencies as appears in \eqref{eq:MultiBand}. Consequently, as previously discussed, any ``off grid'' frequency content in $x(t)$ will spread across the DFT frequencies when the signal is sampled and time-limited.

In the simplified case of a baseband signal model (where $J = 1$, $F_0 = 0$, and $T_s \ll \frac{1}{B_{\textup{nyq}}}$), an efficient alternative to the DFT basis is given by the dictionary of Discrete Prolate Spheroidal Sequences (DPSS's)~\cite{Slepian78DPSS}. DPSS's are a collection of bandlimited sequences that are most concentrated in time to a given index range and the DPSS vectors are the finite-support sequences (or vectors) whose DTFT is most concentrated in a given bandwidth~\cite{Slepian78DPSS}; we review properties of DPSS's in Section~\ref{section:DPSS introduction}. DPSS's provide a highly efficient basis for representing sampled bandlimited signals (when $\mathbb{W}$ reduces to a simple band $[-W_0,W_0]$) and have proved to be useful in numerous signal processing applications. For instance, extrapolating a signal from a finite set of samples is an important problem with applications in remote sensing and other areas~\cite{papoulis1975bandlimitedExtrap}. One can apply DPSS's to find the minimum energy, infinite-length bandlimited sequence that extrapolates a given finite vector of samples~\cite{Slepian78DPSS}. Another problem involves estimating time-varying channels in wireless communication systems. In~\cite{zemen2005channelEstim}, Zemen and Mecklenbr\"{a}uke showed that expressing the time-varying subcarrier coefficients with a DPSS basis yields better estimates than those obtained with a DFT basis, which suffers from frequency leakage.

By modulating the baseband DPSS vectors to different frequency bands and then concatenating these dictionaries, one can construct a new dictionary that provides an efficient representation of sampled multiband signals. Sejdi\'{c} et al.~\cite{SejdicICASSP2008ChannelEstimationDPSS} proposed one such dictionary to provide a sparse representation for fading channels and improve channel estimation accuracy. Zemen et al.~\cite{zemen2007minimum,zemen2012adaptive} utilized multiband DPSS sequences for band-limited prediction and estimation of time-variant channels. In CS, Davenport and Wakin~\cite{DavenportWakin2012CSDPSS} studied multiband modulated DPSS dictionaries for recovery of sampled multiband signals, and Sejdi\'{c} et al.~\cite{sejdic2012compressive} applied these dictionaries for the recovery of physiological signals from compressive measurements. Ahmad et al.~\cite{AhmadQianAmin2015WallCluterDPSS} used such dictionaries for mitigating wall clutter in through-the-wall radar imaging, and modulated DPSS's can also be useful for detecting targets behind the wall~\cite{Zhu2015targetDetectDPSS,Zhu2016targetDetectDPSS}.

In most of these works, the dictionary is assembled by partitioning the digital bandwidth $[-\frac{1}{2},\frac{1}{2}]$ uniformly into many bands and constructing a modulated DPSS basis for each band. The key fact that makes such a dictionary useful is that finite-length sample vectors arising from multiband analog signals will tend to have a block-sparse representation in this dictionary, where only those bands in the dictionary overlapping the frequencies $\mathbb{W}$ are utilized. With this block-sparse structure,~\cite{DavenportWakin2012CSDPSS} provided theoretical guarantees into the use of this dictionary for sparsely representing sampled multiband signals and recovering sampled multiband signals from compressive measurements. For example, using a block-based CS reconstruction algorithm, we are guaranteed that most finite-length sample vectors arising from multiband analog signals can be highly accurately recovered from a number of compressive measurements that is proportional to the multiband signal's total spectral support~\cite[Theorem 5.6]{DavenportWakin2012CSDPSS}.
Experiments demonstrate that reconstruction using the multiband modulated DPSS dictionary yields superior performance compared to reconstruction with the DFT basis.

To date, however, relatively little work has focused on providing formal approximation guarantees for sampled multiband signals using multiband modulated DPSS dictionaries. To the best of our knowledge, an approximation guarantee in a mean-square error (MSE) sense was first presented formally in~\cite{DavenportWakin2012CSDPSS}. However, the question of how this dictionary compares to an optimal one has not been addressed. The objective of this paper is to answer this question and related ones.

\subsection{Contributions}
\label{sec:subspaceArrpxMulit}

We study multiband modulated DPSS dictionaries in terms of the subspaces they span on the respective bands. More specifically, let
\[
\bm{e}_f:=\left[\begin{array}{c}e^{j2\pi f0}\\e^{j2\pi f1}\\\vdots\\e^{j2\pi f(N-1)} \end{array}\right] \in \mathbb{C}^N, ~ f\in[-\frac{1}{2},\frac{1}{2}]
\]
denote a length-$N$ vector of samples from a discrete-time complex exponential signal with digital frequency $f$. Then, it follows directly from \eqref{eq:xsynthesis} that a multiband sample vector $\bm{x}$ can be expressed as
\begin{equation}
\bm{x} = \int_{\mathbb{W}} \widetilde{x}(f) \bm{e}_f \; df,
\label{eq:xvecsynthesis}
\end{equation}
where $\mathbb{W}$ is as defined in \eqref{eq:MultiBand}. We can interpret this equation geometrically: the sampled complex exponentials $\{\bm{e}_f\}_{f \in [-1/2,1/2]}$ comprise a one-dimensional submanifold of $\mathbb{C}^N$. The vectors
\[
\mathcal{M}_\mathbb{W}:=\{\bm{e}_f\}_{f\in\mathbb{W}}
\]
trace out a union of $J$ finite-length curves belonging to this manifold. The sample vector $\bm{x}$ can be expressed as an integral over the vectors in $\mathcal{M}_\mathbb{W}$, with weights determined by $\widetilde{x}(f)$.

We are interested in several questions relating to the union of curves $\mathcal{M}_\mathbb{W}$:
\begin{itemize}
\item What is its effective dimensionality? That is, what dimensionality of a union of subspaces could nearly capture the energy of all signals in $\mathcal{M}_\mathbb{W}$, in the $\ell_2$ metric?
\item What is a suitable basis for the collective span of this union of subspaces?
\end{itemize}
Since we consider $\ell_2$ approximation error, we will approach the approximation problem via the Karhunen-Lo\`{e}ve (KL) transform (also known as principal component analysis (PCA) \cite{jolliffe2002principalCA})~\cite{stark1986probability,DavenportWakin2012CSDPSS}. We can imagine drawing a vector randomly from $\mathcal{M}_\mathbb{W}$ with random phase, and we study the covariance structure of this random vector. Its covariance matrix is $\bm{B}_{N,\mathbb{W}}$, which has entries
\e
\bm{B}_{N,\mathbb{W}}[m,n] := \int_\mathbb{W}e^{j2\pi f(m-n)}df =\sum_{i=0}^{J-1}e^{j2\pi f_i(m-n)}\frac{\sin\left(2\pi W_i(m-n)\right)}{\pi(m-n)}
\label{eq:Bcov}
\ee
for all $m,n\in[N]$. The eigen-decomposition of $\bm{B}_{N,\mathbb{W}}$ provides the optimal dictionary for linearly approximating this random vector. In particular, the $k$ eigenvectors corresponding to the $k$ largest eigenvalues of $\bm{B}_{N,\mathbb{W}}$ span the $k$-dimensional subspace of $\mathbb{C}^N$ that best captures these random vectors in terms of MSE; the resulting MSE equals the sum of the $N-k$ smallest eigenvalues. When $k$ can be chosen such that this residual sum is indeed small, this indicates that the effective dimensionality (informally, the ``information level'') of the set $\mathcal{M}_\mathbb{W}$ is roughly $k$.

The first contribution of this paper is to investigate the spectrum of the matrix $\bm{B}_{N,\mathbb{W}}$, which is equivalent\footnote{By equivalent, we mean that $\bm{B}_{N,\mathbb{W}}\bm x$ = $\mathcal{I}_N(\mathcal{B}_\mathbb{W}(\mathcal{I}^*_N(\bm x)))$ for any $\bm x\in \mathbb{C}^N$.} to a composed time- and multiband- limiting operator $\mathcal{I}_N\mathcal{B}_\mathbb{W}\mathcal{I}^*_N$ defined in Section~\ref{section:Time frequecny limiting operator}. In line with analogous results for time-frequency localization in the continuous-time domain~\cite{Izu2009TimeFrequencyLocalization,landau1993density}, we extend some of the techniques from \cite{Izu2009TimeFrequencyLocalization,landau1993density} for the discrete-time case and show that the number of dominant eigenvalues of $\mathcal{I}_N\mathcal{B}_\mathbb{W}\mathcal{I}^*_N$ (and hence $\bm{B}_{N,\mathbb{W}}$) is essentially the time-frequency area $N|\mathbb{W}| = \sum_i 2NW_i$, which also reveals the effective dimensionality of the union of curves $\mathcal{M}_\mathbb{W}$. Furthermore, similar to the concentration behavior of the DPSS eigenvalues for a single frequency band, we show that the eigenvalues of the operator $\mathcal{I}_N\mathcal{B}_\mathbb{W}\mathcal{I}^*_N$ have a distinctive behavior: the first $\approx N|\mathbb{W}|$ eigenvalues tend to cluster near $1$, while the remaining eigenvalues tend to cluster near $0$ after a narrow transition, which has width proportional to the number of bands times $\log(N)$.  All of these facts tell us that $\approx N|\mathbb{W}|$ atoms are needed in order to accurately approximate, in an MSE sense, discrete-time sinusoids with frequencies in $\mathbb{W}$. As indicated in~\eqref{eq:xvecsynthesis}, such discrete-time sinusoids are themselves the building blocks of sampled multiband signals.

The second contribution of this paper is to show that the multiband modulated DPSS dictionary is approximately the optimal one for representing sampled multiband signals. Specifically, we show that there is a near nesting relationship between the subspaces spanned by the true eigenvectors of $\bm{B}_{N,\mathbb{W}}$ and by the multiband modulated DPSS vectors on the bands of interest.\footnote{By ``bands of interest,'' we mean the union of intervals $\mathbb{F}$ for continuous-time signals and $\mathbb{W}$ for discrete-time signals. We assume these bands are known and are used to construct the multiband modulated DPSS dictionary. The results in this paper, however, can also have application in the problem of detecting the active bands from a set of possible candidates, as was studied in~\cite{DavenportWakin2012CSDPSS}.} Directly computing both baseband DPSS vectors and the eigenvectors of $\bm{B}_{N,\mathbb{W}}$ can be difficult, as the clustering of the eigenvalues makes the problem ill-conditioned. However, several references such as~\cite{Grunbaum1981ToeplitzCommuteTridiaognal,Slepian78DPSS} have pointed out that the baseband DPSS's can also be computed by noting that the corresponding prolate matrix commutes with a well-conditioned symmetric tridiagonal matrix. Thus, the multiband modulated DPSS dictionary, which merely consists of various modulations of baseband DPSS's, can be constructed more easily than the optimal one (which consists of the eigenvectors of $\bm{B}_{N,\mathbb{W}}$).

The third contribution of this paper is to confirm that the multiband modulated DPSS dictionary provides a high degree of approximation for {\it all} sample vectors $\bm{e}_f$ of discrete-time sinusoids with frequencies $f$ in our bands of interest. We also show that for any continuous-time multiband signal that is also approximately time-limited, the resulting finite-length sample vector can be well-approximated by the multiband modulated DPSS dictionary. This result serves as a supplement to~\cite{DavenportWakin2012CSDPSS}, which shows this approximation guarantee is available for a time-limited signal which has its spectrum concentrated in the bands of interest.

We hope that these results will prove useful in the continued study and application of multiband modulated DPSS dictionaries.

The remainder of this paper is organized as follows. Section \ref{section:Prelminaries} defines the time- and multiband-limiting operator and provides some important background information on DPSS's. We state our main results in Section \ref{Section:MainResult}. We conclude in Section \ref{Section:conclusion} with a final discussion.

\section{Preliminaries}\label{section:Prelminaries}

\subsection{Notation}

Finite-dimensional vectors and matrices are indicated by bold characters. We index all such vectors and matrices beginning at $0$. The Hermitian transpose of a matrix $\bm{A}$ is denoted by $\bm{A}^H$. For any natural number $N$, we let $[N]$ denote the set $\{0,1,\ldots,N-1\}$. For any $k \in \{1,2,\dots,N\}$, let $[\bm{A}]_k$ denote the $N \times k$ matrix formed by taking the first $k$ columns of $\bm{A}\in\mathbb{C}^{N\times N}$. In addition, $x(N) \sim y(N)$ means $x$ and $y$ are asymptotically equal, that is $x(N) = y(N) + o(y(N)) = (1+o(1))y(N)$ as $N\rightarrow \infty$.

\subsection{Time, index, and multiband-limiting operators}\label{section:Time frequecny limiting operator}

To begin, let $\mathcal{B}_{\mathbb{W}}: \ell_2(\mathbb{Z})\rightarrow \ell_2(\mathbb{Z})$ denote the multiband-limiting operator that bandlimits the DTFT of a discrete-time signal to the frequency range $\mathbb{W}\subset [-\frac{1}{2},\frac{1}{2}]$, i.e., for $y\in \ell_2(\mathbb{Z})$, we have that
\e
\mathcal{B}_{\mathbb{W}}(y)[m]: = \int_\mathbb{W}e^{j2\pi fm}df\star y[m]=\sum_{n=-\infty}^{\infty}\left(y[n]\int_\mathbb{W}e^{j2\pi f(m-n)}df \right),
\nonumber\ee
where $\star$ stands for convolution.
In addition, let $\mathcal{T}_N: \ell_2(\mathbb{Z})\rightarrow \ell_2(\mathbb{Z})$ denote the operator that zeros out all entries outside the index range $\{0,1,\dots,N-1\}$. That is
\e
\mathcal{T}_N(y)[m]:= \left\{\begin{array}{ll}y[m], & m \in [N],\\0, & \text{otherwise}. \end{array}\right.
\nonumber\ee
Next, define the index-limiting operator $\mathcal{I}_N: \ell_2(\mathbb{Z})\rightarrow \mathbb{C}^N$ as
\e
 \mathcal{I}_N(y)[m]: = y[m], ~ m \in [N].
\nonumber\ee
The adjoint operator $\mathcal{I}^*_N: \mathbb{C}^N\rightarrow \ell_2(\mathbb{Z})$ (anti-index-limiting operator) is given by
\e
\mathcal{I}^*_N(\bm{y})[m]: = \left\{ \begin{array}{ll} \bm{y}[m], & m \in [N],\\
0, & \text{otherwise}. \end{array}\right.
\nonumber\ee
We can observe that $\mathcal{T}_N = \mathcal{I}_N^*\mathcal{I}_N$.

Now the time- and multiband-limiting operator $\mathcal{B}_{\mathbb{W}}\mathcal{T}_N:\ell_2(\mathbb{Z})\rightarrow \ell_2(\mathbb{Z})$ is defined by
\e
\mathcal{B}_{\mathbb{W}}(\mathcal{T}_N(y))[m]: = \sum_{n=0}^{N-1}\left(y[n]\int_\mathbb{W}e^{j2\pi f(m-n)}df \right),~ m \in \mathbb{Z}.
\label{equ-TimeBandLimitingOper}
\ee
Further composing the time- and multiband-limiting operators, we obtain the linear operator $\mathcal{T}_N\mathcal{B}_{\mathbb{W}}\mathcal{T}_N:\ell_2(\mathbb{Z})\rightarrow \ell_2(\mathbb{Z})$ as
\e\begin{split}
\mathcal{T}_N(\mathcal{B}_{\mathbb{W}}(\mathcal{T}_N(y)))[m] = \left\{ \begin{array}{ll} \sum_{n=0}^{N-1}\left(y[n]\int_\mathbb{W}e^{j2\pi f(m-n)}df \right), & m\in[N],\\
0, & \text{otherwise}. \end{array}\right.
\label{equ-TimeBandTimeLimitingOper}\end{split}\ee
Similarly, combining the index- and multiband-limiting operators, we obtain the linear operator $\mathcal{I}_N\mathcal{B}_{\mathbb{W}}\mathcal{I}^*_N:\mathbb{C}^N\rightarrow \mathbb{C}^N$ as
\e
\mathcal{I}_N(\mathcal{B}_{\mathbb{W}}(\mathcal{I}^*_N(\bm{y})))[m] = \sum_{n=0}^{N-1}\left(\bm{y}[n]\int_\mathbb{W}e^{j2\pi f(m-n)}df \right),~ m\in[N].
\label{equ-IndexBandLimitingOper}\ee

Suppose $y' \in \ell_2(\mathbb{Z})$ is an eigenfunction of $\mathcal{T}_N\mathcal{B}_{\mathbb{W}}\mathcal{T}_N$ with corresponding eigenvalue $\lambda'$: $\mathcal{T}_N(\mathcal{B}_{\mathbb{W}}(\mathcal{T}_N(y'))) = \lambda'y'$. We can verify that $\mathcal{I}_N(\mathcal{B}_{\mathbb{W}}(\mathcal{I}^*_N(\mathcal{I}_N(y')))) = \lambda'\mathcal{I}_N(y')$. On the other hand, if $\bm{y}''$ and $\lambda''$ satisfy $\mathcal{I}_N(\mathcal{B}_{\mathbb{W}}(\mathcal{I}^*_N(\bm{y}''))) = \lambda''\bm{y}''$, then we can conclude that $\mathcal{T}_N(\mathcal{B}_{\mathbb{W}}(\mathcal{T}_N(\mathcal{I}_N^*(\bm{y}'')))) =  \lambda''\mathcal{I}_N^*(\bm{y}'')$. Therefore  $\mathcal{T}_N\mathcal{B}_{\mathbb{W}}\mathcal{T}_N$ and $\mathcal{I}_N\mathcal{B}_{\mathbb{W}}\mathcal{I}^*_N$ have the same eigenvalues, and the eigenvectors of $\mathcal{I}_N\mathcal{B}_{\mathbb{W}}\mathcal{I}^*_N$ can be obtained by index-limiting the eigenvectors of $\mathcal{T}_N\mathcal{B}_{\mathbb{W}}\mathcal{T}_N$.

Note that $\mathcal{I}_N\mathcal{B}_{\mathbb{W}}\mathcal{I}^*_N$ is equivalent to the covariance matrix $\bm{B}_{N,\mathbb{W}}$ (see~\eqref{eq:Bcov}), as a linear operator on $\mathbb{C}^N$. Thus, in order to answer the questions raised in Section~\ref{sec:subspaceArrpxMulit}, we will study the eigenvalue concentration behavior of $\mathcal{I}_N\mathcal{B}_\mathbb{W}\mathcal{I}^*_N$.

\subsection{DPSS bases for sampled bandlimited signals}\label{section:DPSS introduction}

In this subsection, we briefly review important definitions and properties of DPSS's from~\cite{DavenportWakin2012CSDPSS,Slepian78DPSS}.

\subsubsection{DPSS's and DPSS vectors}

\begin{defn}(DPSS's \cite{Slepian78DPSS}) Given $W\in(0,\frac{1}{2})$ and $N\in\mathbb{N}$, the Discrete Prolate Spheroidal Sequences (DPSS's) $\{s_{N,W}^{(0)},s_{N,W}^{(1)},\dots,s_{N,W}^{(N-1)}\}$ are real-valued discrete-time sequences that satisfy $\mathcal{B}_{[-W,W]}(\mathcal{T}_N(s_{N,W}^{(l)}))=\lambda_{N,W}^{(l)}s_{N,W}^{(l)}$ for all $l\in[N]$. Here $\lambda_{N,W}^{(0)},\dots,\lambda_{N,W}^{(N-1)}$ are the eigenvalues of the operator $\mathcal{B}_{[-W,W]}\mathcal{T}_N$ with order $1>\lambda_{N,W}^{(0)}>\lambda_{N,W}^{(1)}>\dots>\lambda_{N,W}^{(N-1)}>0$.
\end{defn}

The DPSS's are orthogonal on $\mathbb{Z}$ and on $\{0,1,\dots,N-1\}$, and they are normalized so that
\e
\langle\mathcal{T}_N(s_{N,W}^{(k)}),\mathcal{T}_N(s_{N,W}^{(l)})\rangle = \begin{cases} 1, & k = l \\ 0, & k \neq l. \end{cases}
\nonumber\ee
Consequently, it can be shown~\cite{Slepian78DPSS} that $||s_{N,W}^{(l)}||_2^2 = (\lambda_{N,W}^{(l)})^{-1}$. The vector obtained by index-limiting $s_{N,W}^{(l)}$ to the index range $\{0,1,\ldots,N-1\}$ is an eigenvector of the $N\times N$ matrix $\bm{B}_{N,W}$ with elements given by\footnote{For convenience, we use $\bm{B}_{N,W}$ instead of $\bm{B}_{N,[-W,W]}$ to denote the matrix which is equivalent to the operator $\mathcal{I}_N\mathcal{B}_{[-W,W]}\mathcal{I}^*_N$. This is also the reason that we use $\lambda_{N,W}$, $s_{N,W}$ and  $\bm{s}_{N,W}$ (which will be defined later) instead of $\lambda_{N,[-W,W]}$, $s_{N,[-W,W]}$ and  $\bm{s}_{N,[-W,W]}$. }
\begin{equation}
\bm{B}_{N,W}[m,n] := \int_{-W}^We^{j2\pi f(m-n)}df=\frac{\sin(2\pi W(m-n))}{\pi(m-n)}.
\nonumber\end{equation}

DPSS's are useful for constructing a dictionary that efficiently represents index-limited versions of sampled bandlimited signals. As pointed out in~\cite{DavenportWakin2012CSDPSS}, the index-limited DPSS's also satisfy $\mathcal{I}_N(\mathcal{B}_{[-W,W]}(\mathcal{T}_N(s_{N,W}^{(l)})))=\lambda_{N,W}^{(l)}\mathcal{I}_N(s_{N,W}^{(l)})$.

\begin{defn}(DPSS vectors \cite{Slepian78DPSS}) Given $W\in(0,\frac{1}{2})$ and $N\in\mathbb{N}$, the DPSS vectors $\bm{s}_{N,W}^{(0)}$ $\bm{s}_{N,W}^{(1)},\dots,~\bm{s}_{N,W}^{(N-1)}\in\mathbb{R}^N$ are defined by index-limiting the DPSS's to the index range $\{0,1,\ldots,N-1\}$:
$$\bm{s}_{N,W}^{(l)}=\mathcal{I}_N(s_{N,W}^{(l)})$$
and satisfy
$$
\mathcal{I}_N(\mathcal{B}_{[-W,W]}(\mathcal{I}^*_N(\bm{s}_{N,W}^{(l)}))) =\bm{B}_{N,W}\bm{s}_{N,W}^{(l)}=\lambda_{N,W}^{(l)}\bm{s}_{N,W}^{(l)}.
$$
\end{defn}

It follows that $\bm{B}_{N,W}$ can be factorized as
$$
\bm{B}_{N,W} = \bm{S}_{N,W}\bm{\Lambda}_{N,W}\bm{S}_{N,W}^H,
$$
where $\bm{\Lambda}_{N,W}$ is an $N\times N$ diagonal matrix whose diagonal elements are the DPSS eigenvalues $\lambda_{N,W}^{(0)},\lambda_{N,W}^{(1)},\ldots,\lambda_{N,W}^{(N-1)}$ and $\bm{S}_{N,W}$ is a square ($N\times N$) matrix whose $l$-th column is the DPSS vector $\bm{s}_{N,W}^{(l)}$ for all $l\in [N]$.

The following provides a useful result on the clustering of the eigenvalues $\lambda^{(0)}_{N,W},\lambda^{(1)}_{N,W},\ldots, \lambda^{(N-1)}_{N,W}$.

\begin{lem}(Clustering of eigenvalues \cite{DavenportWakin2012CSDPSS,Slepian78DPSS}) Suppose that $W\in(0,\frac{1}{2})$ is fixed.
\begin{enumerate}
\item Fix $\epsilon\in(0,1)$. Then there exist constants $C_1(W,\epsilon),C_2(W,\epsilon)$ (which may depend on $W,\epsilon$) and an integer $N_0(W,\epsilon)$ (which may also depend on $W,\epsilon$) such that
\begin{equation}
1-\lambda^{(l)}_{N,W}\leq C_1(W,\epsilon)e^{-C_2(W,\epsilon)N}, ~\forall ~ l\leq\lfloor2NW(1-\epsilon)\rfloor
\nonumber\end{equation}
for all $N\geq N_0(W,\epsilon)$.
\item Fix $\epsilon\in(0,\frac{1}{2W}-1)$. Then there exist constants $C_3(W,\epsilon),C_4(W,\epsilon)$ (which may depend on $W,\epsilon$) and an integer $N_1(W,\epsilon)$ (which may also depend on $W,\epsilon$) such that
\begin{equation}
\lambda^{(l)}_{N,W}\leq C_3(W,\epsilon)e^{-C_4(W,\epsilon)N}, ~ \forall ~ l\geq\lceil2NW(1+\epsilon)\rceil
\nonumber\end{equation}
for all $N\geq N_1(W,\epsilon)$.
\end{enumerate}
\label{lem:EigClusteringDPSS}\end{lem}

In words, the first $\approx 2NW$ eigenvalues tend to cluster very close to $1$, while the remaining eigenvalues tend to cluster very close to $0$. As a consequence of this behavior, the effective dimensionality of the vectors $\mathcal{M}_{[-W,W]}:=\{\bm{e}_f\}_{f\in[-W,W]}$ (which trace out a finite-length curve in $\mathbb{C}^N$) is essentially $2NW$, in the sense that we can use a subspace formed by the first $\approx 2NW$ DPSS vectors to approximate this curve with low MSE.

\subsubsection{DPSS bases for sampled bandpass signals}

Let us now consider the eigenvectors of the operator $\mathcal{I}_N(\mathcal{B}_{[f_c-W,f_c+W]}(\mathcal{I}^*_N))$, which can be expressed as:
\begin{equation}
\begin{split}
\mathcal{I}_N(\mathcal{B}_{[f_c-W,f_c+W]}(\mathcal{I}^*_N(\bm{y})))[m]=&\left(\int_{f_c-W}^{f_c+W}e^{j2\pi m f}df\right)\star \left(\mathcal{I}^*_N(\bm{y})[m]\right)\\
=&\sum_{n=0}^{N-1}e^{j2\pi f_c(m-n)}\frac{\sin\left(2\pi W(m-n)\right)}{\pi (m-n)}\bm{y}[n]
\end{split}
\nonumber\end{equation}
for all $m = 0,1,\ldots,N-1$. Let $\bm{E}_{f_c}$ denote an $N\times N$ diagonal matrix with entries
\e
\bm{E}_{f_c}[m,n]:=\left\{\begin{array}{ll}e^{j2\pi f_c m}, & m=n,\\0, & m \neq n.\end{array}\right.
\nonumber\ee
We can verify that the modulated DPSS vectors $\bm{E}_{f_c}\bm{s}_{N,W}^{(0)},\bm{E}_{f_c}\bm{s}_{N,W}^{(1)},\dots,\bm{E}_{f_c}\bm{s}_{N,W}^{(N-1)}$ satisfy
$$
\mathcal{I}_N(\mathcal{B}_{[f_c-W,f_c+W]}(\mathcal{I}^*_N(\bm{E}_{f_c}\bm{s}_{N,W}^{(l)})))=\bm{E}_{f_c}\bm{B}_{N,W}\bm{E}_{f_c}^H\bm{E}_{f_c}\bm{s}_{N,W}^{(l)}=\lambda_{N,W}^{(l)}\bm{E}_{f_c}\bm{s}_{N,W}^{(l)}.
$$
That is, $(\lambda_{N,W}^{(l)},\bm{E}_{f_c}\bm{s}_{N,W}^{(l)})$ is an \textit{eigenpair} of the operator $\mathcal{I}_N(\mathcal{B}_{[f_c-W,f_c+W]}(\mathcal{I}^*_N))$ for all $l\in[N]$.

For any integer $k\in\{1,2,\ldots,N\}$, let $\bm{Q} :=[\bm{E}_{f_c}\bm{S}_{N,W}]_k$ denote the $N\times k$ matrix formed by taking the first $k$ modulated DPSS vectors. Also let $\bm{P}_{\bm Q}$ denote the orthogonal projection onto the column space of $\bm Q$. It is shown in \cite{DavenportWakin2012CSDPSS} that the dictionary $\bm{Q}$ provides very accurate approximations (in an MSE sense) for finite-length sample vectors arising from sampling random bandpass signals.

\begin{thm}(\cite{DavenportWakin2012CSDPSS} Theorem 4.2) Suppose $x$ is a continuous, zero-mean, wide sense stationary random process with power spectrum
$$P_{x}(F)=\left\{\begin{array}{ll} \frac{1}{B_{\textup{band}}},& F\in[F_c-\frac{B_{\textup{band}}}{2},F_c+\frac{B_{\textup{band}}}{2}], \\0, & \textup{otherwise}.\end{array}\right.$$
Let $\bm{x} = [x(0) ~x(T_s) ~ \ldots ~x((N-1)T_s)]^T\in \mathbb{C}^N$ denote a finite vector of samples acquired from $x(t)$ with a sampling interval of $T_s\leq 1/(2\max\{|F_c\pm \frac{B_\textup{band}}{2}|\})$.
 Let $f_c = F_cT_s$ and $W = \frac{B_{\textup{band}}T_s}{2}$. We will have
 $$
 \mathbb{E}\left[||\bm{x}-\bm{P}_{\bm{Q}}\bm{x}||_2^2\right] = \frac{1}{2W}\int_{f_c-W}^{f_c+W}||\bm{e}_f-\bm{P}_{\bm{Q}}\bm{e}_f||_2^2df =  \frac{1}{2W}\sum_{l=k}^{N-1}\lambda_{N,W}^{(l)}.
 $$
Furthermore, for fixed $\epsilon\in(0,\frac{1}{2W}-1)$, set $k = 2NW(1+\epsilon)$. Then
\e
\mathbb{E}\left[||\bm{x}-\bm{P}_{\bm{Q}}\bm{x}||_2^2\right] \leq \frac{C_3(W,\epsilon)}{2W}Ne^{-C_4(W,\epsilon)N}\ee
for all $N\geq N_1(W,\epsilon)$, where $N_1(W,\epsilon)$, $C_3(W,\epsilon)$, $C_4(W,\epsilon)$ are constants specified in Lemma \ref{lem:EigClusteringDPSS}. For comparison, $\mathbb{E}\left[||\bm{x}||_2^2\right] = ||\bm{e}_f||_2^2 = N$.
\label{thm:DPSSApproxBandpass}\end{thm}

\section{Main Results}\label{Section:MainResult}

We now consider the multiband case,  where 
\e
\mathbb{W} = [f_0-W_0,f_0+W_0]\cup[f_1-W_1,f_1+W_1]\cup\cdots\cup[f_{J-1}-W_{J-1},f_{J-1}+W_{J-1}] \subseteq \left[-\frac{1}{2},\frac{1}{2}\right]
\nonumber\ee
is a union of $J$ intervals as in~\eqref{eq:MultiBand}. For each $i\in[J]$, define $\bm{\Psi}_i = [\bm{E}_{f_i}\bm{S}_{N,W_i}]_{k_i}$ for some value $k_i\in\{1,2,\ldots,N\}$ that we can choose as desired. We construct the multiband modulated DPSS dictionary $\bm{\Psi}$ by concatenating these subdictionaries:
\e
\bm{\Psi}:=[\bm{\Psi}_0 ~ \bm{\Psi}_1 ~ \cdots ~ \bm{\Psi}_{J-1}].
\label{def_mul_dpss_dic}\ee
In this section, we investigate the efficiency of using $\bm{\Psi}$ to represent discrete-time sinusoids and sampled multiband signals.

\subsection{Eigenvalues for time- and multiband-limiting operator}
\label{sec:opteig}

We begin by studying the eigenvalue concentration behavior of the operator $\mathcal{I}_N\mathcal{B}_\mathbb{W}\mathcal{I}^*_N$ (and hence $\bm{B}_{N,\mathbb{W}}$), which reveals the effective dimensionality of the finite union of curves $\mathcal{M}_{\mathbb{W}}=\{\bm{e}_f\}_{f\in\mathbb{W}}$.

We first establish the following rough bound, which states that all the eigenvalues of $\mathcal{I}_N\mathcal{B}_\mathbb{W}\mathcal{I}^*_N$ are between $0$ and $1$.

\begin{lem}
For any $\mathbb{W}\subset[-\frac{1}{2},\frac{1}{2}]$ and $N$, the operator $\mathcal{I}_N\mathcal{B}_{\mathbb{W}}\mathcal{I}^*_N$ is positive-definite with eigenvalues
$$
1>\lambda_{N,\mathbb{W}}^{(0)}\geq\lambda_{N,\mathbb{W}}^{(1)}\geq\cdots\geq\lambda_{N,\mathbb{W}}^{(N-1)}>0
$$ and
$$\sum_{l=0}^{N-1}\lambda_{N,\mathbb{W}}^{(l)} = N|\mathbb{W}|.$$
We denote the corresponding eigenvectors of $\mathcal{I}_N\mathcal{B}_{\mathbb{W}}\mathcal{I}^*_N$ by $\bm{u}_{N,\mathbb{W}}^{(0)},\bm{u}_{N,\mathbb{W}}^{(1)},\ldots,\bm{u}_{N,\mathbb{W}}^{(N-1)}$.
\label{lem:PSD}\end{lem}

\vspace{0.3cm}
\noindent{\textbf{Proof.}} See Appendix \ref{proof:PSD}.
\vspace{0.3cm}

There is, in fact, a sharp transition in the distribution of the eigenvalues of $\mathcal{I}_N\mathcal{B}_\mathbb{W}\mathcal{I}^*_N$. We establish this fact in the following theorem.

\begin{thm}
Suppose $\mathbb{W}$ is a finite union of $J$ pairwise disjoint intervals as defined in (\ref{eq:MultiBand}). For any $\varepsilon\in(0,\frac{1}{2})$, the number of eigenvalues of $\mathcal{I}_N\mathcal{B}_\mathbb{W}\mathcal{I}^*_N$ that are between $\varepsilon$ and $1-\varepsilon$ satisfies
\e
\#\{l:\varepsilon\leq\lambda^{(l)}_{N,\mathbb{W}}\leq 1-\varepsilon\} \leq J\frac{\frac{2}{\pi^2}\log(N-1)+\frac{2}{\pi^2}\frac{2N-1}{N-1}}{\varepsilon(1-\varepsilon)}.
\label{Eq_num_eigenvalue_transition}\ee
\label{num_eigenvalue_transition}\end{thm}

\vspace{0.3cm}
\noindent{\textbf{Proof.}} See Appendix \ref{proof_num_eigenvalue_transition}.
\vspace{0.3cm}

This result states that the number of eigenvalues in $[\epsilon,1-\epsilon]$ is in the order of $\log(N)$ for any fixed $\epsilon\in(0,\frac{1}{2})$. Along with the following result which states that the number of eigenvalues of $\mathcal{I}_N\mathcal{B}_\mathbb{W}\mathcal{I}^*_N$ greater than $\frac{1}{2}$ equals  $\approx N|\mathbb{W}|$, we conclude that the effective dimensionality of $\mathcal{M}_{\mathbb{W}}$ is approximately $N|\mathbb{W}| = \sum_i 2NW_i$.

\begin{thm}
Let $\mathbb{W}\subset[-\frac{1}{2},\frac{1}{2}]$ be a finite union of $J$ disjoint intervals having the form in (\ref{eq:MultiBand}). Denote by 
$$\iota_- = \#\{n\in\mathbb{Z}:-\lfloor\frac{N}{2}\rfloor\leq n\leq\lfloor\frac{N-1}{2}\rfloor,~(\frac{n}{N}-\frac{1}{2N},\frac{n}{N}+\frac{1}{2N})\subset\mathbb{W}\}\label{eq_lminus}$$
and
$$\iota_+ = \#\{n\in\mathbb{Z}:-\lfloor\frac{N}{2}\rfloor\leq n\leq\lfloor\frac{N-1}{2}\rfloor,~(\frac{n}{N}-\frac{1}{2N},\frac{n}{N}+\frac{1}{2N})\cap\mathbb{W}\neq\emptyset\}.$$
In particular, it holds that $\lfloor N|\mathbb{W}|\rfloor-2J+2\leq\iota_-\leq\iota_+\leq\lceil N|\mathbb{W}|\rceil+2J-2$. Then the eigenvalues of the operator $\mathcal{I}_N\mathcal{B}_\mathbb{W}\mathcal{I}^*_N$ satisfy
$$
\lambda_{N,\mathbb{W}}^{(\iota_--1)}\geq\frac{1}{2}\geq\lambda_{N,\mathbb{W}}^{(\iota_+)}.
$$
\label{thm_eigenvalues_greater_half}\end{thm}
\vspace{0.3cm}
\noindent{\textbf{Proof.}} See Appendix \ref{proof_thm_eigenvalues_greater_half}.
\vspace{0.3cm}

Note that results similar to the above two theorems for time-frequency localization in the continuous domain have been established in~\cite{hogan2011bookDurationBandLimiting,Izu2009TimeFrequencyLocalization,landau1993density}. Similar to the ideas used in~\cite{hogan2011bookDurationBandLimiting}, the key to proving Theorem~\ref{num_eigenvalue_transition} is to obtain an upper bound on the distance between the trace of $\mathcal{I}_N\mathcal{B}_\mathbb{W}\mathcal{I}^*_N$ and the sum of the squared eigenvalues of $\mathcal{I}_N\mathcal{B}_\mathbb{W}\mathcal{I}^*_N$. Constructing an appropriate subspace with a carefully selected bandlimited sequence for the Weyl-Courant minimax characterization of eigenvalues is the key to proving Theorem~\ref{thm_eigenvalues_greater_half}. The proof techniques of~\cite{Izu2009TimeFrequencyLocalization,landau1993density} form the basis of our analysis in Appendix~\ref{proof_thm_eigenvalues_greater_half}, but some modifications are required to extend their results to the discrete domain.

Similar to what happens in the single band case (when $J=1$; see Lemma~\ref{lem:EigClusteringDPSS}), the eigenvalues of $\mathcal{I}_N\mathcal{B}_\mathbb{W}\mathcal{I}^*_N$ have a distinctive behavior: the first $N|\mathbb{W}| = \sum_i 2NW_i$ eigenvalues tend to cluster very close to $1$, while the remaining eigenvalues tend to cluster very close to 0, after a narrow transition. This is captured formally in the following result.

\begin{thm}
Let $\mathbb{W}\subset[-\frac{1}{2},\frac{1}{2}]$ be a fixed finite union of $J$ disjoint intervals having the form in (\ref{eq:MultiBand}).
\begin{enumerate}
\item Fix $\epsilon\in(0,1)$. Then there exist constants $\overline{C}_1(\mathbb{W},\epsilon),\overline{C}_2(\mathbb{W},\epsilon)$ (which may depend on $\mathbb{W}$ and $\epsilon$) and an integer $\overline{N}_0(\mathbb{W},\epsilon)$ (which may also depend on $\mathbb{W}$ and $\epsilon$) such that
$$
\lambda^{(l)}_{N,\mathbb{W}}\geq1-\overline{C}_1(\mathbb{W},\epsilon)N^2e^{-\overline{C}_2(\mathbb{W},\epsilon)N}, ~\forall~ l\leq J-1+\sum_i \lfloor 2NW_i(1-\epsilon)\rfloor
$$
for all $N\geq \overline{N}_0(\mathbb{W},\epsilon)$.
\item Fix $\epsilon\in(0,\frac{1}{|\mathbb{W}|}-1)$. Then there exist constants $\overline{C}_3(\mathbb{W},\epsilon),\overline{C}_4(\mathbb{W},\epsilon)$ (which may depend on $\mathbb{W}$ and $\epsilon$) and an integer $\overline{N}_1(\mathbb{W},\epsilon)$ (which may also depend on $\mathbb{W}$ and $\epsilon$) such that
$$
\lambda^{(l)}_{N,\mathbb{W}}\leq\overline{C}_3(\mathbb{W},\epsilon)e^{-\overline{C}_4(\mathbb{W},\epsilon)N}, ~\forall~ l\geq \sum_i \lceil2NW_i(1+\epsilon)\rceil
$$
for all $N\geq \overline{N}_1(\mathbb{W},\epsilon)$.
\end{enumerate}

We point out that $\overline{N}_0(\mathbb{W},\epsilon)\geq \max{\{N_{0}(W_i,\epsilon), ~\forall~ i \in[J]\}}$, $\overline{C}_2(\mathbb{W},\epsilon) = \frac{\min{\{C_{2}(W_i,\epsilon),~\forall~i \in [J]\}}}{2}$, $\overline{C}_3(\mathbb{W},\epsilon)  = J\max{\{C_{3}(W_i,\epsilon), ~\forall~ i \in[J]\}}$ and $\overline{C}_4(\mathbb{W},\epsilon) = \min{\{C_{4}(W_i,\epsilon), ~\forall~ i \in[J]\}}$, which will prove useful in our analysis below. Here $C_2(W_i,\epsilon)$, $C_3(W_i,\epsilon)$, and $C_4(W_i,\epsilon)$ are as specified in Lemma~\ref{lem:EigClusteringDPSS}.
\label{thm:EigConcentrationMulitband}\end{thm}
\vspace{0.3cm}
\noindent{\textbf{Proof.}} See Appendix \ref{proof:EigConcentrationMulitband}.
\vspace{0.3cm}

\subsection{Multiband modulated DPSS dictionaries for sampled multiband signals}

Let $p\in \{1,2,\ldots,N\}$. Define
\e
\bm{\Phi} := [\bm{u}_{N,\mathbb{W}}^{(0)} ~ \bm{u}_{N,\mathbb{W}}^{(1)} ~ \cdots ~ \bm{u}_{N,\mathbb{W}}^{(p-1)}],
\label{def_Phi}\ee
where $\bm{u}_{N,\mathbb{W}}^{(l)}, ~\forall~ l\in[N]$ are the eigenvectors of $\mathcal{I}_N\mathcal{B}_{\mathbb{W}}\mathcal{I}^*_N$.
Let $\bm{\Psi}$ be the multiband modulated DPSS dictionary defined in (\ref{def_mul_dpss_dic}).

There are three main reasons why the dictionary $\bm{\Psi}$ may be useful representing sampled multiband signals. First, direct computation of $\bm{\Phi}$ is difficult due to the clustering of the eigenvalues of $\bm{B}_{N,\mathbb{W}}$. However, in the single band case, the matrix $\bm{B}_{N,W}$ is known to commute with a symmetric tridiagonal matrix that has well-separated eigenvalues, and hence its eigenvectors can be efficiently and stably computed~\cite{Slepian78DPSS}. Gr\"{u}nbaum \cite{Grunbaum1981ToeplitzCommuteTridiaognal} gave a certain condition for a Toeplitz matrix to commute with a tridiagonal matrix with a simple spectrum. We can check that the matrix $\bm{B}_{N,\mathbb{W}}$ in general does not satisfy this condition, except for the case when $\mathbb{W}$ consists of only a single interval. However, we emphasize that $\bm{\Psi}$ is constructed simply by modulating DPSS's, which, again, can be computed efficiently.

Second, the multiband modulated DPSS dictionary $\bm{\Psi}$ provides an efficient representation for sampled multiband signals. Davenport and Wakin~\cite{DavenportWakin2012CSDPSS} provided theoretical guarantees into the use of this dictionary for sparsely representing sampled multiband signals and recovering sampled multiband signals from compressive measurements. We extend one of these guarantees in Section~\ref{sec:approxmb}. Moreover, we confirm that a multiband modulated DPSS dictionary provides a high degree of approximation for all discrete-time sinusoids with frequencies in $\mathbb{W}$ in Section~\ref{sec_approx_pure_tone_gen}.

Third, as indicated by the results in Section~\ref{sec:opteig}, $\approx N|\mathbb{W}|$ dictionary atoms are necessary in order to achieve a high degree of approximation for the discrete-time sinusoids in an MSE sense. Our results, along with~\cite{DavenportWakin2012CSDPSS}, show that the multiband modulated DPSS dictionary $\bm{\Psi}$ with $\approx N|\mathbb{W}|$ atoms can indeed approximate discrete-time sinusoids with high accuracy. In order to help explain this result, we first show that there is a near nesting relationship between the subspaces spanned by the columns of $\bm{\Psi}$ and by the columns of the optimal dictionary $\bm{\Phi}$.

\subsubsection{The subspace angle between $\mathcal{S}_{\bm{\Psi}}$ and $\mathcal{S}_{\bm{\Phi}}$}

In order to compare subspaces of possibly different dimensions, we require the following definition of angle between subspaces.

\begin{defn}
Let $\mathcal{S}_{\bm{\Psi}}$ and $\mathcal{S}_{\bm{\Phi}}$ be the subspaces formed by the columns of the matrices $\bm{\Psi}$ and $\bm{\Phi}$ respectively. The subspace angle $\Theta_{\mathcal{S}_{\bm{\Psi}}\mathcal{S}_{\bm{\Phi}}}$ between $\mathcal{S}_{\bm{\Psi}}$ and $\mathcal{S}_{\bm{\Phi}}$ is given by
$$
\cos(\Theta_{\mathcal{S}_{\bm{\Psi}}\mathcal{S}_{\bm{\Phi}}}):=\inf_{\bm{\phi}\in\mathcal{S}_{\bm{\Phi}}, ||\bm{\phi}||_2=1}||\bm{P}_{\bm{\Psi}}\bm{\phi}||_2
$$
if $dim(\mathcal{S}_{\bm{\Psi}})\geq dim(\mathcal{S}_{\bm{\Phi}})$, or
$$
\cos(\Theta_{\mathcal{S}_{\bm{\Psi}}\mathcal{S}_{\bm{\Phi}}}):=\inf_{\bm{\psi}\in\mathcal{S}_{\bm{\Psi}}, ||\bm{\psi}||_2=1}||\bm{P}_{\bm{\Phi}}\bm{\psi}||_2
$$
if $dim(\mathcal{S}_{\bm{\Psi}})<dim(\mathcal{S}_{\bm{\Phi}})$. Here $\bm{P}_{\bm \Psi}$ (or $\bm{P}_{\bm \Phi}$) denotes the orthogonal projection onto the column space of $\bm \Psi$ (or $\bm \Phi$).
\end{defn}

Our first guarantee considers the case where in constructing $\bm{\Psi}$, each $k_i$ is chosen slightly smaller than $2NW_i$, and in constructing $\bm{\Phi}$, we take $p$ to be slightly larger than $\sum_i 2NW_i$. In this case, we can guarantee that the subspace angle between $\mathcal{S}_{\bm{\Psi}}$ and $\mathcal{S}_{\bm{\Phi}}$ is small.

\begin{thm} Let $\mathbb{W}\subset[-\frac{1}{2},\frac{1}{2}]$ be a fixed finite union of $J$ disjoint intervals having the form in (\ref{eq:MultiBand}). Fix $\epsilon\in(0,\min{\{1,\frac{1}{|\mathbb{W}|}-1\}})$. Let $p=\sum_i \lceil2NW_i(1+\epsilon)\rceil$ and $\bm{\Phi}$ be the $N\times p$ matrix defined in (\ref{def_Phi}). Also let $k_i\leq\lfloor 2NW_i(1-\epsilon)\rfloor, \forall i\in[J]$ and $\bm{\Psi}$ be the matrix defined in (\ref{def_mul_dpss_dic}). Then for any column $\bm{\psi}$ in $\bm{\Psi}$,
$$
||\bm{\psi}-\bm{P}_{\bm{\Phi}}\bm{\psi}||_2^2\leq \frac{2\widetilde{C}_1(\mathbb{W},\epsilon)e^{-\widetilde{C}_2(\mathbb{W},\epsilon)N}}{\left(1-\widetilde{C}_1(\mathbb{W},\epsilon)e^{-\widetilde{C}_2(\mathbb{W},\epsilon)N}-\overline{C}_3(\mathbb{W},\epsilon)e^{-\overline{C}_4(\mathbb{W},\epsilon)N}\right)^2} =: \kappa_1(N,\mathbb W,\epsilon)
$$
and
\e
\cos(\Theta_{\mathcal{S}_{\bm{\Psi}}\mathcal{S}_{\bm{\Phi}}}) \geq \sqrt{\frac{1 - \kappa_1(N,\mathbb W,\epsilon)-N\sqrt{\kappa_1(N,\mathbb W,\epsilon)} - 3N\sqrt{\widetilde{C}_1(\mathbb{W},\epsilon)}e^{-\frac{\widetilde{C}_2(\mathbb{W},\epsilon)}{2}N}}{ 1+ 3N \sqrt{\widetilde{C}_1(\mathbb{W},\epsilon)}e^{-\frac{\widetilde{C}_2(\mathbb{W},\epsilon)}{2}N}}
}
\label{eq_subsapce_angle_1_2}\ee
if $N\geq \max\{\overline{N}_0(\mathbb{W},\epsilon),\overline{N}_1(\mathbb{W},\epsilon)\}$. Here $\widetilde{C}_1(\mathbb{W},\epsilon) = \max{\{C_{1}(W_i,\epsilon), ~\forall~ i \in[J]\}}$, $\widetilde{C}_2(\mathbb{W},\epsilon) = \min{\{C_{2}(W_i,\epsilon),~\forall~i \in [J]\}}$, $\overline{N}_0(\mathbb{W},\epsilon)$, $\overline{N}_1(\mathbb{W},\epsilon)$, $\overline{C}_3(\mathbb{W},\epsilon)$, and $\overline{C}_4(\mathbb{W},\epsilon)$ are the constants specified in Theorem~\ref{thm:EigConcentrationMulitband}, and $C_1(W_i,\epsilon)$ and $C_2(W_i,\epsilon)$ are the constants specified in Lemma~\ref{lem:EigClusteringDPSS}.
\label{thm_subspace_angle_1}\end{thm}

\vspace{0.3cm}
\noindent{\textbf{Proof.}} See Appendix \ref{proof_thm_subspace_angle_1}.
\vspace{0.3cm}

We can also guarantee that the subspace angle between $\mathcal{S}_{\bm{\Psi}}$ and $\mathcal{S}_{\bm{\Phi}}$ is small if, in constructing $\bm{\Psi}$, each $k_i$ is chosen slightly larger than $2NW_i$, and in constructing $\bm{\Phi}$, we take $p$ to be slightly smaller than $\sum_i 2NW_i$. This result is established in Corollary~\ref{cor_subspace_approx_3}, which follows from Theorem~\ref{thm_subspace_approx_2}.

\begin{thm} Let $\mathbb{W}\subset[-\frac{1}{2},\frac{1}{2}]$ be a finite union of $J$ disjoint intervals having the form in (\ref{eq:MultiBand}). Given some values $k_i\in\{1,2,\ldots,N\}, \forall i\in[J]$, let $\bm{\Psi}$ be the matrix defined in (\ref{def_mul_dpss_dic}). Then
$$
||\bm{P}_{\bm{\Psi}}\bm{u}_{N,\mathbb{W}}^{(l)}||_2\geq \lambda_{N,\mathbb{W}}^{(l)} - \sum_{i=0}^{J-1}\sum_{l_i=k_i}^{N-1}\lambda_{N,W_i}^{(l_i)}
$$
for all $l\in\{0,1,\ldots,N-1\}$.
\label{thm_subspace_approx_2}\end{thm}

\vspace{0.3cm}
\noindent{\textbf{Proof.}} See Appendix \ref{proof_thm_subspace_approx_2}.
\vspace{0.3cm}

\begin{cor} Let $\mathbb{W}\subset[-\frac{1}{2},\frac{1}{2}]$ be a fixed finite union of $J$ disjoint intervals having the form in (\ref{eq:MultiBand}). Fix $\epsilon\in(0,\min\{1,\frac{1}{|\mathbb{W}|}-1\})$. Let $p\leq J-1+\sum_i \lfloor2NW_i(1-\epsilon)\rfloor$ and $\bm{\Phi}$ be the $N\times p$ matrix defined in (\ref{def_Phi}). Also let $k_i=\lceil 2NW_i(1+\epsilon)\rceil, \forall i\in[J]$ and $\bm{\Psi}$ be the matrix defined in (\ref{def_mul_dpss_dic}). Then for any column $\bm{u}_{N,\mathbb{W}}^{(l)}$ in $\bm{\Phi}$,
$$
||\bm{P}_{\bm{\Psi}}\bm{u}_{N,\mathbb{W}}^{(l)}||_2\geq 1 - \overline{C}_1(\mathbb{W},\epsilon)N^2e^{-\overline{C}_2(\mathbb{W},\epsilon)N} - N\overline{C}_3(\mathbb{W},\epsilon)e^{-\overline{C}_4(\mathbb{W},\epsilon)N}
$$
and
\e
\cos(\Theta_{\mathcal{S}_{\bm{\Psi}}\mathcal{S}_{\bm{\Phi}}}) \geq \sqrt{1 - 2\kappa_2(N,\mathbb W,\epsilon) + \kappa_2^2(N,\mathbb W,\epsilon) - N\sqrt{2\kappa_2(N,\mathbb W,\epsilon) - \kappa_2^2(N,\mathbb W,\epsilon)}}\label{eq:subsapceAngle2-2}\ee
for all $N\geq \max\{\overline{N}_0(\mathbb{W},\epsilon),\overline{N}_1(\mathbb{W},\epsilon)\}$,
where $\overline{N}_0(\mathbb{W},\epsilon), \overline{N}_1(\mathbb{W},\epsilon), \overline{C}_1(\mathbb{W},\epsilon), \overline{C}_2(\mathbb{W},\epsilon), \overline{C}_3(\mathbb{W},\epsilon)$ and $\overline{C}_4(\mathbb{W},\epsilon)$ are constants specified in Theorem~\ref{thm:EigConcentrationMulitband}, and $\kappa_2(N,\mathbb W,\epsilon) $ is defined as $ \kappa_2(N,\mathbb W,\epsilon) := \overline{C}_1(\mathbb{W},\epsilon)N^2e^{-\overline{C}_2(\mathbb{W},\epsilon)N}+ N\overline{C}_3(\mathbb{W},\epsilon)e^{-\overline{C}_4(\mathbb{W},\epsilon)N}$.
\label{cor_subspace_approx_3}\end{cor}

\vspace{0.3cm}
\noindent{\textbf{Proof.}} See Appendix~\ref{proof_cor_subspace_approx_3}.
\vspace{0.3cm}

Although our results hold for scenarios where one dictionary contains $\sum_i \lfloor2NW_i(1-\epsilon)\rfloor$ atoms while another one has $\sum_i \lceil2NW_i(1+\epsilon)\rceil$ atoms, we note that these dimensions can be made very close by choosing $\epsilon$ sufficiently small.\footnote{Though a small $\epsilon$ may require $N$ large enough such that our results hold, $\frac{\sum_i \lfloor2NW_i(1-\epsilon)\rfloor}{\sum_i \lceil2NW_i(1+\epsilon)\rceil}$ (the ratio between the sizes of the two dictionaries) may become close to $1$.}

\subsubsection{Approximation quality for discrete-time sinusoids}
\label{sec_approx_pure_tone_gen}

The above results show that $\bm{\Psi}$ spans nearly the same space as $\bm{\Phi}$ in the case where both dictionaries contain $\approx N|\mathbb{W}|$ columns. In this section, we investigate the approximation quality of $\bm{\Psi}$ for discrete-time sinusoids with frequencies in the bands of interest. Then, in the next section, we investigate the approximation quality of $\bm{\Psi}$ for sampled multiband signals.

We first prove that a single band dictionary with slightly more than $2NW$ baseband DPSS vectors can capture almost all of the energy in any sinusoid with a frequency in $[-W,W]$. Our analysis is based upon an expression for the DTFT of the DPSS vectors proposed in~\cite{Slepian78DPSS}. We review this result in Appendix~\ref{sec:DTFTofDPSS}.

\begin{thm}
Fix $W\in(0,\frac{1}{2})$ and $\epsilon\in(0,\frac{1}{2W}-1)$. Let $W' = \frac{1}{2} - W$, $\epsilon' = \frac{W}{\frac{1}{2}-W}\epsilon$ and $k=2NW(1+\epsilon)$. Then there exists a constant $C_9(W',\epsilon')$ (which may depend on $W'$ and $\epsilon'$) such that
$$
||\bm{e}_f - \bm{P}_{[\bm{S}_{N,W}]_k}\bm{e}_f||_2^2 \leq C_9(W',\epsilon')N^{5/2}e^{-C_2(W',\epsilon')N}, ~ \forall |f|\leq W
$$
for all $N\geq N_0(W',\epsilon')$, where $N_0(W',\epsilon')$ and $C_2(W',\epsilon')$ are constants defined in Lemma \ref{lem:EigClusteringDPSS}.
\label{thm:DPSSApproxPureTone}\end{thm}
\vspace{0.3cm}
\noindent{\textbf{Proof.}}  See Appendix \ref{proof:DPSSApproxPureTone}.
\vspace{0.3cm}

To the best of our knowledge, this is the first work that rigorously shows that {\em every} discrete-time sinusoid with a frequency $f \in [-W,W]$ is well-approximated by a DPSS basis $[\bm{S}_{N,W}]_k$ with $k$ slightly larger than $2NW$. This result extends the approximation guarantee in an MSE sense presented in~\cite{DavenportWakin2012CSDPSS}. We now extend this result for the multiband modulated DPSS dictionary.

\begin{cor}
Let $\mathbb{W}\subset[-\frac{1}{2},\frac{1}{2}]$ be a fixed finite union of $J$ disjoint intervals having the form in (\ref{eq:MultiBand}). Fix $\epsilon\in(0,\frac{1}{|\mathbb{W}|}-1)$. Let $k_i=2NW_i(1+\epsilon), \forall i\in[J]$ and $\bm{\Psi}$ be the matrix defined in (\ref{def_mul_dpss_dic}). Then there exist constants $C_{10}(\mathbb{W},\epsilon)$ and $C_{11}(\mathbb{W},\epsilon)$ (which may depend on $\mathbb{W}$ and $\epsilon$) and an integer $N_2(\mathbb{W},\epsilon)$ (which may also depend on $\mathbb{W}$ and $\epsilon$) such that
\e
||\bm{e}_f - \bm{P}_{\bm{\Psi}}\bm{e}_f||_2^2 \leq C_{10}(\mathbb{W},\epsilon)N^{5/2}e^{-{C}_{11}(\mathbb{W},\epsilon)N}, ~ \forall f\in \mathbb{W}
\label{eq:MDPSSApproxPureTone}\ee
for all $N\geq N_2(\mathbb{W},\epsilon)$.
\label{cor:MDPSSApproxPureTone}\end{cor}
\vspace{0.3cm}
\noindent{\textbf{Proof.}}  See Appendix~\ref{proof:corMDPSSApproxPureTone}.
\vspace{0.3cm}

\subsubsection{Approximation quality for sampled multiband signals (statistical analysis)}
\label{sec:approxmb}

As indicated in~\cite{DavenportWakin2012CSDPSS}, in a probabilistic sense, most finite-length sample vectors arising from multiband analog signals can be well-approximated by the multiband modulated DPSS dictionary. In this final section, we generalize the result~\cite[Theorem 4.4]{DavenportWakin2012CSDPSS} to sampled multiband signals where each band has a possibly different width.

\begin{thm}
Suppose for each $i\in[J]$, $x_i(t)$ is a continuous-time, zero-mean, wide sense stationary random process with power spectrum
\e P_{x_i}(F)=\left\{\begin{array}{ll} \frac{1}{\sum_{i=0}^{J-1}B_{\textup{band}_i}},& F_i - \frac{B_{\textup{band}_i}}{2}\leq F \leq F_i + \frac{B_{\textup{band}_i}}{2} \\0, & \text{otherwise},\end{array}\right.,\label{eq:powerspectrumMD}\ee
and furthermore suppose $x_0(t), x_1(t), \ldots, x_{J-1}(t)$ are independent and jointly wide sense stationary. Let $T_s$ denote a sampling interval chosen to satisfy the minimum Nyquist sampling rate, which means $T_s \leq \frac{1}{B_{\textup{nyq}}}: = 1/\left(2\max\left\{\left|F_i\pm \frac{B_{\textup{band}_i}}{2}\right|,~\forall~i\in[J]\right\}\right)$. Let $\bm{x}_i = [x_i(0) ~x_i(T_s) ~ \ldots ~x_i((N-1)T_s)]^T\in \mathbb{C}^N$ denote a finite vector of samples acquired from $x_i(t)$ and let $\bm{x} = \sum_{i=1}^J \bm{x}_i$. Set $f_i = F_iT_s$ and $W_i = \frac{B_{\textup{band}_i}T_s}{2}$. Let $\bm{\Psi}$ be the matrix defined in (\ref{def_mul_dpss_dic}) for some given $k_i$. Then
\begin{equation}
\mathbb{E}[||\bm{x}-\bm{P}_{\bm{\Psi}}\bm{x}||_2^2]\leq \frac{1}{|\mathbb{W}|}\sum_{i=0}^{J-1}\sum_{l_i=k_i}^{N-1}\lambda_{N,W_i}^{(l_i)},
\label{eq:mbprob}
\end{equation}
where $\mathbb{E}[||\bm{x}||_2^2]=N$.
\label{thm_appr_uniform}
\end{thm}

\vspace{0.3cm}
\noindent{\textbf{Proof.}} See Appendix \ref{proof_thm_appr_uniform}.~
\vspace{0.3cm}

The right hand side of~\eqref{eq:mbprob} can be made small by choosing $k_i \approx 2NW_i$ for each $i \in [J]$; recall Lemma~\ref{lem:EigClusteringDPSS}. Aside from allowing for different band widths, the above result improves the upper bound of~\cite[Theorem 4.4]{DavenportWakin2012CSDPSS} by a factor of $J$.

Finally, the following result establishes a deterministic guarantee for the approximation of sampled multiband signals using a multiband modulated DPSS dictionary with $\approx N|\mathbb{W}|$ atoms.

\begin{cor}
Suppose $x$ is a continuous-time signal with Fourier transform $X(F)$ supported on $\mathbb{F} = \mathop{\cup}\limits_{i=0}\limits^{J-1}[F_i-B_{\text{band}_i}/2,F_i+B_{\text{band}_i}/2]$, i.e., $$
x(t) = \int_{\mathbb{F}}X(F)e^{j2\pi Ft}dF.
$$

Let $\bm{x} = [x(0) ~x(T_s) ~ \ldots ~x((N-1)T_s)]^T\in \mathbb{C}^N$ denote a finite vector of samples acquired from $x(t)$ with a sampling interval of $T_s\leq 1/(2\max\{|F_c\pm \frac{B_\textup{band}}{2}|\})$. Let $W_i = T_sB_{\text{band}_i}/2$, $f_i = T_sF_i$ for all $i\in[J]$, and $\mathbb{W} = \mathop{\cup}\limits_{i=0}\limits^{J-1} [f_i-W_i,f_i+W_i]$.
 Fix $\epsilon\in(0,\frac{1}{|\mathbb{W}|}-1)$. Let $k_i=2NW_i(1+\epsilon), \forall i\in[J]$ and let $\bm{\Psi}$ be the matrix defined in (\ref{def_mul_dpss_dic}). Then
\e
||\bm{x}-\bm{P}_{\bm{\Psi}}\bm{x}||_2^2\leq \left(\int_{\mathbb{W}}|\widetilde{x}(f)|^2 \;df \right) \cdot  C_{10}(\mathbb{W},\epsilon)N^{5/2}e^{-{C}_{11}(\mathbb{W},\epsilon)N}
\label{eq:MDPSSApproxMultiband}\ee
for all $N\geq N_2(\mathbb{W},\epsilon)$, where $N_2(\mathbb{W},\epsilon)$, $C_{10}(\mathbb{W},\epsilon)$ and $C_{11}(\mathbb{W},\epsilon)$ are constants specified in Corollary~\ref{cor:MDPSSApproxPureTone}.
\label{cor:MDPSSApproxMultiband}
\end{cor}
\vspace{0.3cm}
\noindent{\textbf{Proof.}}  See Appendix~\ref{proof:corMDPSSApproxMultiband}.
\vspace{0.3cm}

Corollary~\ref{cor:MDPSSApproxMultiband} can be applied in various settings:
\begin{itemize}
\item The sequence $x[n]$ encountered in most practical problems has finite energy. For example, if we assume that $\int_{\mathbb{W}}|\widetilde{x}(f)|^2df\leq 1$, we conclude that $||\bm{x}-\bm{P}_{\bm{\Psi}}\bm{x}||_2^2\leq C_{10}(\mathbb{W},\epsilon)N^{5/2}e^{-{C}_{11}(\mathbb{W},\epsilon)N}$.
\item Moreover, in some practical problems, the finite-energy sequence $x[n]$ may be approximately time-limited to the index range $n = 0,1,\ldots, N-1$ such that for some $\delta$, $||\bm{x}||_2^2=||\mathcal{I}_N(x)||_2^2\geq (1-\delta)||x||_2^2$. In this case, (\ref{eq:MDPSSApproxMultiband}) guarantees that
\e\frac{||\bm{x}-\bm{P}_{\bm{\Psi}}\bm{x}||_2^2}{||\bm{x}||_2^2}\leq\frac{\int_{\mathbb{W}}|\widetilde{x}(f)|^2df}{||\bm{x}||_2^2}\cdot C_{10}(\mathbb{W},\epsilon)N^{5/2}e^{-{C}_{11}(\mathbb{W},\epsilon)N}\leq \frac{1}{1-\delta}C_{10}(\mathbb{W},\epsilon)N^{5/2}e^{-{C}_{11}(\mathbb{W},\epsilon)N},\label{eq:determGuarantee}\ee
where the last inequality follows from Parseval's theorem that $||x||_2^2 = \int_\mathbb{W}|\widetilde{x}(f)|^2df$.
\end{itemize}

Along with the result proved in~\cite{DavenportWakin2012CSDPSS} that samples from a time-limited sequence which is approximately bandlimited to the bands of interest can be well-approximated by the multiband modulated DPSS dictionary, we conclude that the multiband modulated DPSS dictionary is useful for most practical problems involving representing sampled multiband signals.

However, we point out that not {\em all} sampled multiband signals can be well-approximated by the multiband modulated DPSS dictionary. To illustrate this, consider the simple case where $\mathbb{W}$ reduces to a single band $[-W,W]$. Recalling that the infinite-length DPSS's are strictly bandlimited, it follows that each of the DPSS vectors can be obtained by sampling and time-limiting some strictly bandlimited analog signal. Nevertheless, for all $l\geq k$, we will have
\begin{equation}
\frac{||\bm{s}_{N,W}^{(l)} - \bm{P}_{[\bm{S}_{N,W}]_{k}}\bm{s}_{N,W}^{(l)} ||_2}{||\bm{s}_{N,W}^{(l)} ||_2} = 1
\label{eq:pessbound}
\end{equation}
even when we choose $k = 2NW(1+\epsilon)$. In this case, the approximation guarantee in~\eqref{eq:pessbound} is much worse than what appears in~\eqref{eq:determGuarantee}. Such examples are pathological, however: the infinite sequence $s_{N,W}^{(l)}$ has energy $||s_{N,W}^{(l)}||_2^2 = (\lambda_{N,W}^{(l)})^{-1}$, which according to Lemma~\ref{lem:EigClusteringDPSS} is exponentially large when $l \ge 2NW(1+\epsilon)$, and yet the energy of the sampled vector $||\bm{s}_{N,W}^{(l)}||_2^2$ is only $1$. Moreover, the spectrum of the infinite sequence $s_{N,W}^{(l)}$ is entirely concentrated in the band $[-W,W]$ while the spectrum of the time-limited sequence $\mathcal{T}_N(s_{N,W}^{(l)})$ is almost entirely contained outside the band $[-W,W]$, and so on.  We refer the reader to~\cite{DavenportWakin2012CSDPSS} for additional discussion of this topic.

\section{Conclusions}
\label{Section:conclusion}

In this paper, we have provided a thorough analysis of the spectrum of a time- and multiband-limiting operator in the discrete-time domain. We have showed that the information level of finite-length multiband sample vectors is essentially equal to their time-frequency area, which also indicates the number of dictionary atoms required in order to obtain a high-quality approximation. We have also considered the angle between the subspaces spanned by the eigenfunctions of the time- and multiband-limiting operator and by the multiband modulated DPSS dictionary. Our results show that the multiband modulated DPSS dictionary is nearly optimal in terms of representing finite-length vectors arising from sampling multiband analog signals.

We have showed that the multiband modulated DPSS dictionary can not only guarantee a very high degree of approximation accuracy in an MSE sense for finite-length multiband sample vectors, but also that it can guarantee such accuracy uniformly over all discrete-time sinusoids in the bands of interest. Though we are not guaranteed such accuracy uniformly over all sampled multiband signals, we have suggested that such accuracy holds for most practical problems involving multiband signals. Thus, our work supports the growing evidence that multiband modulated DPSS dictionaries can be useful for engineering applications.

\section*{Acknowledgements}

We gratefully acknowledge Mark Davenport, Armin Eftekhari, and Justin Romberg for valuable discussions and insightful comments; and the anonymous reviewers for their constructive comments.

\appendix

\section{Proof of Lemma \ref{lem:PSD}}\label{proof:PSD}

\noindent\textbf{Proof.} Let $\bm{y}\in\mathbb{C}^N,\bm{y}\neq \bm{0}$ be an arbitrary vector. Then
\begin{equation}\begin{split}
\langle\mathcal{I}_N(\mathcal{B}_{\mathbb{W}}(\mathcal{I}^*_N(\bm{y}))),\bm{y}\rangle &=\sum_{m=0}^{N-1}\mathcal{I}_N(\mathcal{B}_{\mathbb{W}}(\mathcal{I}^*_N(\bm{y})))[m]\overline{\bm{y}}[m]=\sum_{m=0}^{N-1}\left(\sum_{n=0}^{N-1}\int_{\mathbb{W}}e^{j2\pi f(m-n)}df\bm{y}[n]\right)\overline{\bm{y}}[m]\\
&=\int_{\mathbb{W}}\left(\sum_{m=0}^{N-1}e^{j2\pi f m}\overline{\bm{y}}[m]\right)\left(\sum_{n=0}^{N-1}e^{-j2\pi f n}\bm{y}[n]\right)df=\int_{\mathbb{W}}|\sum_{n=0}^{N-1}\bm{y}[n]e^{-j2\pi f n}|^2df>0,
\end{split}\nonumber\end{equation}
where $\overline{\bm{y}}$ is the complex-conjugate of the vector $\bm{y}$, $\sum_{n=0}^{N-1}\bm{y}[n]e^{-j2\pi f n}$ is the DTFT of $\mathcal{I}^*_N(\bm{y})$, and the last inequality is derived from the fact that compactly supported signals cannot have perfectly flat magnitude response.

By Parsevel's Theorem, we know $\int_{-1/2}^{1/2}|\sum_{n=0}^{N-1}\bm{y}[n]e^{-j2\pi f n}|^2d f=||\bm{y}||_2^2$. Therefore
\begin{equation}
\langle\mathcal{I}_N(\mathcal{B}_{\mathbb{W}}(\mathcal{I}^*_N(\bm{y}))),\bm{y}\rangle=\int_{\mathbb{W}}|\sum_{n=0}^{N-1}\bm{y}[n]e^{-j2\pi f n}|^2d f<||\bm{y}||_2^2.
\nonumber\end{equation}
Thus, we have
\e
0<\min_{\bm{y}\in \mathbb{C}^N} \frac{\langle\mathcal{I}_N(\mathcal{B}_{\mathbb{W}}(\mathcal{I}^*_N(\bm{y}))),\bm{y}\rangle}{||\bm{y}||_2^2}\leq\lambda_{N,\mathbb{W}}^{(l)}\leq\max_{\bm{y}\in \mathbb{C}^N} \frac{\langle\mathcal{I}_N(\mathcal{B}_{\mathbb{W}}(\mathcal{I}^*_N(\bm{y}))),\bm{y}\rangle}{||\bm{y}||_2^2}<1
\nonumber\ee
for all $l\in[N]$.

By noting that  $\mathcal{I}_N\mathcal{B}_{\mathbb{W}}\mathcal{I}^*_N$ is equivalent to  $\bm{B}_{N,\mathbb{W}}$, we have
$$
\sum_{l=0}^{N-1}\lambda_{N,\mathbb{W}}^{(l)} = \text{trace}(\bm{B}_{N,\mathbb{W}})=\sum_{n=0}^{N-1}\bm{B}_{N,\mathbb{W}}[n,n]=\sum_{n=0}^{N-1}\int_{\mathbb{W}}e^{j2\pi f 0}df=N|\mathbb{W}|. ~~~\square
$$

\section{Proof of Theorem \ref{num_eigenvalue_transition}}\label{proof_num_eigenvalue_transition}

\noindent\textbf{Proof.} First we state a useful inequality about the Frobenius norm of positive semi-definite matrices.
Suppose $\bm{X}\in\mathbb{C}^{N\times N}$ and $\bm{Y}\in\mathbb{C}^{N\times N}$ are two arbitrary positive semi-definite matrices. Then
\e\begin{split}
||\bm{X}+\bm{Y}||_F^2 &= \textup{trace}{\left((\bm{X}+\bm{Y})^H(\bm{X}+\bm{Y})\right)}\\ &=||\bm{X}||_F^2+||\bm{Y}||_F^2 +2 \textup{trace}{(\bm{X}^H\bm{Y})}\\
&\geq||\bm{X}||_F^2+||\bm{Y}||_F^2,
\end{split}\nonumber\ee
where the last inequality is derived from the fact that $\textup{trace}(\bm{X}^H\bm{Y})$ is nonnegative, which can be showed as follows. By the hypothesis that
 $\bm{X}$ and $\bm{Y}$ are positive semi-definite matrices, we have the factorization $\bm{X}^H = \bm{X} = \bm{X}^{1/2}\bm{X}^{1/2}$, where $\bm{X}^{1/2}$ is also a positive semi-definite matrix.\footnote{Note that $\bm X$ has the eigen-decomposition $\bm{X} = \bm{V}\bm{D}\bm{V}^H$ where $\bm V$ is an orthonormal matrix and $\bm D$ is a diagonal matrix whose diagonal elements are non-negative, giving the square root $\bm{X}^{1/2} = \bm V \bm{D}^{1/2}\bm V^H$.} Then we conclude that $\text{trace}{(\bm{X}^H\bm{Y})} = \text{trace}{(\bm{X}^{1/2}\bm{X}^{1/2}\bm{Y})} = \text{trace}{(\bm{X}^{1/2}\bm{Y}\bm{X}^{1/2})}\geq 0$, since $\bm{X}^{1/2}\bm{Y}\bm{X}^{1/2}$ is also a positive semi-definite matrix.

We next bound the Frobenius norm of $\bm{B}_{N,W_i}$ by
\e\begin{split}
||\bm{B}_{N,W_i}||_F^2 &= N(2W_i)^2+\underset{m\neq n}{\sum\sum}\left(\frac{\sin\left(2\pi W_i(m-n)\right)}{\pi(m-n)}\right)^2\\
&=4NW_i^2+2\sum_{n=1}^{N-1}(N-n)\left(\frac{\sin\left(2\pi W_in\right)}{\pi n}\right)^2\\
& = 4NW_i^2+2N\sum_{n=1}^{N-1}\left(\frac{\sin\left(2\pi W_in\right)}{\pi n}\right)^2-2\sum_{n=1}^{N-1}n\left(\frac{\sin\left(2\pi W_in\right)}{\pi n}\right)^2\\
& = 4NW_i^2+2N\left(W_i-2W_i^2-\sum_{n=N}^{\infty}\left(\frac{\sin\left(2\pi W_in\right)}{\pi n}\right)^2\right)-2\sum_{n=1}^{N-1}n\left(\frac{\sin\left(2\pi W_in\right)}{\pi n}\right)^2\\
& \geq 4NW_i^2+2N\left(W_i-2W_i^2-\frac{1}{\pi^2}\int_{N-1}^\infty \frac{1}{x^2}dx\right)-2\frac{1}{\pi^2}\left(\int_{1}^{N-1}\frac{1}{x}dx+1\right)\\
& = 2NW_i-\frac{2}{\pi^2}\frac{2N-1}{N-1}-\frac{2}{\pi^2}\log(N-1),
\end{split}\nonumber\ee
where the fourth line follows from Parseval's theorem $\sum_{n=-\infty}^{\infty}\left(\frac{\sin\left(2\pi W_in\right)}{\pi n}\right)^2 = \int_{-W_i}^{W_i}df = 2W_i$, which indicates that $\sum_{n=1}^{\infty}\left(\frac{\sin\left(2\pi W_in\right)}{\pi n}\right)^2 = W_i-2W_i^2$.

Now applying the above results yields
\e\begin{split}
||\bm{B}_{N,\mathbb{W}}||_F^2 & = ||\sum_{i=0}^{J-1}\bm{E}_{f_i}\bm{B}_{N,W_i}\bm{E}_{f_i}^H||_F^2\\
&\geq \sum_{i=0}^{J-1}||\bm{B}_{N,W_i}||_F^2\\
&\geq \sum_{i=0}^{J-1}\left( 2NW_i-\frac{2}{\pi^2}\frac{2N-1}{N-1}-\frac{2}{\pi^2}\log(N-1)\right)\\
& = N|\mathbb{W}|-J\left(\frac{2}{\pi^2}\frac{2N-1}{N-1}+\frac{2}{\pi^2}\log(N-1)\right),
\end{split}\nonumber\ee
where the second line follows since $\bm{E}_{f_i}\bm{B}_{N,W_i}\bm{E}_{f_i}^H$ is positive semi-definite.
Recalling the result stated in Lemma \ref{lem:PSD} that $\sum_{l=0}^{N-1}\lambda_{N,\mathbb{W}}^{(l)}  =\text{trace}(\bm{B}_{N,\mathbb{W}}) = N|\mathbb{W}|$, we get
\e
\sum_{l=0}^{N-1}\lambda_{N,\mathbb{W}}^{(l)}(1-\lambda_{N,\mathbb{W}}^{(l)})= \text{trace}(\bm{B}_{N,\mathbb{W}}) - ||\bm{B}_{N,\mathbb{W}}||_F^2\leq J\left(\frac{2}{\pi^2}\frac{2N-1}{N-1}+\frac{2}{\pi^2}\log(N-1)\right).
\nonumber\ee
Thus, equation (\ref{Eq_num_eigenvalue_transition}) follows by noting that for any $\varepsilon\in(0,\frac{1}{2})$ one has
\e
\sum_{l=0}^{N-1}\lambda_{N,\mathbb{W}}^{(l)}(1-\lambda_{N,\mathbb{W}}^{(l)}) \geq \sum_{\{l:\varepsilon\leq\lambda_{N,\mathbb{W}}^{(l)}\leq 1-\varepsilon\}}\lambda_{N,\mathbb{W}}^{(l)}(1-\lambda_{N,\mathbb{W}}^{(l)})\geq \varepsilon(1-\varepsilon)\#\{l:\varepsilon\leq\lambda_{N,\mathbb{W}}^{(l)}\leq 1-\varepsilon\}. ~~~\square
\nonumber\ee

\section{Proof of Theorem \ref{thm_eigenvalues_greater_half}}\label{proof_thm_eigenvalues_greater_half}

\noindent\textbf{Proof.} A precise proof of a similar result for time- and band-limiting operators in the continuous domain was first given in~\cite{landau1993density}. Izu and Lakey~\cite{Izu2009TimeFrequencyLocalization} extend the result to multiple intervals in the frequency domain or time domain. Their work forms the foundation of the following analysis.

As we have noted, the two operators $\mathcal{T}_N\mathcal{B}_{\mathbb{W}}\mathcal{T}_N$ and $\mathcal{I}_N\mathcal{B}_{\mathbb{W}}\mathcal{I}^*_N$ have the same eigenvalues. We work with $\mathcal{T}_N\mathcal{B}_{\mathbb{W}}\mathcal{T}_N$ to prove Theorem \ref{thm_eigenvalues_greater_half}. For convenience, we also use $\lambda_{N,\mathbb{W}}^{(0)},\lambda_{N,\mathbb{W}}^{(1)},\ldots,\lambda_{N,\mathbb{W}}^{(N-1)}$ to denote the decreasing eigenvalues for the operator $\mathcal{T}_N\mathcal{B}_{\mathbb{W}}\mathcal{T}_N$. We let $S([N])$ denote the subspace of all finite-energy sequences supported only on the index set $[N]$, that is
$$S([N]) = \{y: y\in \ell_2(\mathbb{Z}), \mathcal{T}_N(y) = y\}.$$

First, for all integers $l\in [N]$, the Weyl-Courant minimax representation of the eigenvalues can be stated as
\e\begin{split}
\lambda_{N,\mathbb{W}}^{(l)} &= \left\{\begin{array}{c}\min_{S_l}\max_{y\in \ell_2(\mathbb{Z}),y\perp S_l}\frac{\langle \mathcal{T}_N(\mathcal{B}_{\mathbb{W}}(\mathcal{T}_N(y))),y\rangle}{\langle y,y\rangle},\\
\max_{S_{l+1}}\min_{y\in \ell_2(\mathbb{Z}),y\in S_{l+1}}\frac{\langle \mathcal{T}_N(\mathcal{B}_{\mathbb{W}}(\mathcal{T}_N(y))),y\rangle}{\langle y,y\rangle},
\end{array}\right.\\
& =
\left\{\begin{array}{c}\min_{S_l}\max_{y\in S([N]),y\perp S_l}\frac{\langle \mathcal{T}_N(\mathcal{B}_{\mathbb{W}}(\mathcal{T}_N(y))),y\rangle}{\langle y,y\rangle},\\
\max_{S_{l+1}}\min_{y\in S([N]),y\in S_{l+1}}\frac{\langle \mathcal{T}_N(\mathcal{B}_{\mathbb{W}}(\mathcal{T}_N(y))),y\rangle}{\langle y,y\rangle},
\end{array}\right.\\
& =
\left\{\begin{array}{c}\min_{S_l}\max_{y\in S([N]),y\perp S_l}\frac{\int_{\mathbb{W}}|\widetilde{y}(f)|^2df}{||y||_2^2},\\
\max_{S_{l+1}}\min_{y\in S([N]),y\in S_{l+1}}\frac{\int_{\mathbb{W}}|\widetilde{y}(f)|^2df}{||y||_2^2},
\end{array}\right.
\end{split}\label{eq_minimax_eigenvalue}\ee
where $S_l$ is an $l$-dimensional subspace of $\ell_2(\mathbb{Z})$, and $\widetilde{y}(f)$ is the DTFT of the sequence $y$. Noting that all the eigenvectors of $\mathcal{T}_N\mathcal{B}_{\mathbb{W}}\mathcal{T}_N$ belong to $S([N])$, we restrict to $y\in S([N])$ in the second line.

\begin{lem}
 Consider the bandlimited sequence $g\in\ell_2(\mathbb{Z})$ whose DTFT is given by
\e
\widetilde{g}(f)=\left\{\begin{array}{cl}\sqrt{2N}\cos(N\pi f)e^{-j2\pi f\lfloor\frac{N}{2}\rfloor}, & |f|\leq \frac{1}{2N},\\
0, & \frac{1}{2N}<|f|\leq \frac{1}{2}.\end{array}\right.
\label{def_subspacefuc_g}\ee
Then $||g||_2^2 = 1$ and $g[n]\geq \frac{1}{\sqrt{2N}}$ for all $n\in[N]$.
\label{lem_subspacefuc_g}\end{lem}
\vspace{0.3cm}
\noindent\textbf{Proof} (of Lemma
\ref{lem_subspacefuc_g}). First it is easy to check that $||g||_2^2 = \int_{-\frac{1}{2}}^{\frac{1}{2}}|\widetilde{g}(f)|^2df = 1$. Then computing the inverse DTFT directly yields
$$g[n] = \frac{1}{\sqrt{2N}}\text{sinc}\left( \frac{n-\lfloor\frac{N}{2}\rfloor}{N}-\frac{1}{2}\right) + \frac{1}{\sqrt{2N}}\text{sinc}\left( \frac{n-\lfloor\frac{N}{2}\rfloor}{N}+\frac{1}{2}\right).$$
Let $\xi(t) = \text{sinc}(t-\frac{1}{2})+\text{sinc}(t+\frac{1}{2})$. Taking the directive of $\xi(t)$, we would find on $[-\frac{1}{2},\frac{1}{2}]$ that $\xi(t)$ achieves its minimum value of $1$ at the points $t=\pm \frac{1}{2}$. Therefore, $g[n]\geq\frac{1}{\sqrt{2N}}$ since $|\frac{n-\lfloor\frac{N}{2}\rfloor}{N}|\leq \frac{1}{2}$  for all $n\in[N]$. ~$\square$
 \subsection{Upper bound}
From equation (\ref{eq_minimax_eigenvalue}), we know that
$$\lambda_{N,\mathbb{W}}^{(l)}=\min_{S_l}\max_{y\in S([N]),y\perp S_l}\frac{\int_{\mathbb{W}}|\widetilde{y}(f)|^2df}{||y||_2^2}.$$
Therefore, in order to bound the eigenvalues from above, it suffices to pick an appropriate $l$-dimensional subspace $S_l\subset\ell_2(\mathbb{Z})$ and then find a uniform upper bound for the quantity above for all time-limited sequences $y\in S([N])$ orthogonal to $S_l$.

 Consider the bandlimited sequence $g\in\ell_2(\mathbb{Z})$ defined in (\ref{def_subspacefuc_g}). Let $\mathcal{E}_{f_0}:\ell_2(\mathbb{Z})\rightarrow \ell_2(\mathbb{Z})$ denote a modulating operator with $\mathcal{E}_{f_0}(y)[n]:=e^{j2\pi f_0n}y[n]$ for all $n\in \mathbb{Z}$ and $f_0\in[-\frac{1}{2},\frac{1}{2}]$.
 Set
  $$L_+ = \{n'\in\mathbb{Z}:-\lfloor\frac{N}{2}\rfloor\leq n'\leq\lfloor\frac{N-1}{2}\rfloor,~(\frac{n'}{N}-\frac{1}{2N},\frac{n'}{N}+\frac{1}{2N})\cap\mathbb{W}\neq\emptyset\}$$
and hence $\iota_+ = \#L_+$. Let $S_{\iota_+}$ be the $\iota_+$-dimensional subspace of $\ell_2(\mathbb{Z})$ spanned by the functions $\mathcal{E}_{\frac{n'}{N}}g, n'\in L_+$, that is,
$$S_{\iota_+}:= \text{span}\left(\{\mathcal{E}_{\frac{n'}{N}}g\}_{ n'\in L_+}\right).$$
If the time-limited sequence $y\in S([N])$ is orthogonal to $S_{\iota_+}$, then
$$0=\langle y,\mathcal{E}_{\frac{n'}{N}}g\rangle=\langle \widetilde{y},\widetilde{g}(\cdot-\frac{n'}{N})\rangle =\left( \widetilde{y}\star \widetilde{\overline{g}}\right)\left( \frac{n'}{N}\right)=:g_y[n'], ~n'\in L_+, $$
where $\overline{g}: = g^*$ is the complex-conjugate of the sequence $g$ and $\widetilde{\overline{g}}$ is the DTFT of $\overline{g}$.

Now it follows that
\e\begin{split}\sum_{n'=-\lfloor\frac{N}{2}\rfloor}^{\lfloor\frac{N-1}{2}\rfloor}|g_y[n']|^2
&=\sum_{n'\in L_+^C}|g_y[n']|^2 \\
&=\sum_{n'\in L_+^C}|\int_{\frac{n'-1/2}{N}}^{\frac{n'+1/2}{N}}\widetilde{y}(f)\widetilde{\overline{g}}(\frac{n'}{N}-f)df|^2\\
&\leq\sum_{n'\in L_+^C}\left(||g||_2^2\int_{\frac{n'-1/2}{N}}^{\frac{n'+1/2}{N}}|\widetilde{y}(f)|^2df\right)\\
&\leq \int_{f\notin\mathbb{W}}|\widetilde{y}(f)|^2df\\
&=||y||_2^2-\int_{f\in\mathbb{W}}|\widetilde{y}(f)|^2df,
\end{split}\label{eq_upperbound_energy_g_1}\ee
where $L_+^C$ is defined as $L_+^C := \{n'\in\mathbb{Z}:-\lfloor\frac{N}{2}\rfloor\leq n'\leq\lfloor\frac{N-1}{2}\rfloor,~n'\notin L_+\}$, the second line holds because $g$ is bandlimited to $[-\frac{1}{2N},\frac{1}{2N}]$, the third line follows from the Cauchy-Schwarz inequality, and the fourth line holds because $||g||_2=1$ and by construction, the set $\cup_{n'\in L_+}[\frac{n'}{N}-\frac{1}{2N},\frac{n'}{N}+\frac{1}{2N}]$ covers the intervals $\mathbb{W}$ completely. On the other hand,
let $y\odot \overline{g}$ denote the pointwise product between $y$ and $\overline{g}$, that is $(y\odot \overline{g})[n] = y[n]\overline{g}[n]$. Note that $y\odot\overline{g}$ has the same support in time as $y$, namely $[N]$, and $\{\frac{1}{\sqrt{N}}{\bm e}_{\frac{n'}{N}},-\lfloor\frac{N}{2}\rfloor\leq n'\leq\lfloor\frac{N-1}{2}\rfloor\}$ forms an orthobasis (normalized DFT basis) for $\mathbb{C}^N$. We can rewrite $g_y[n'] = \bm{e}_{\frac{n'}{N}}^H\left(y\odot\overline{g}\right)$, which can be viewed as the DFT of $y\odot\overline{g}$. Therefore, using Parseval's theorem, we acquire
\e
\sum_{n'=-\lfloor\frac{N}{2}\rfloor}^{\lfloor\frac{N-1}{2}\rfloor}|g_y[n']|^2 = N||y\odot\overline{g}||_2^2\geq \frac{1}{2}||y||_2^2
\nonumber\ee
since by hypothesis, $g[n]\geq \frac{1}{\sqrt{2N}}$ for all $n\in[N]$.
Now, combining the above lower bound on the energy of the sequence $g_y$ and the upper bound in (\ref{eq_upperbound_energy_g_1}), we observe that
$$\frac{1}{2}||y||_2^2\leq ||y||_2^2-\int_{f\in\mathbb{W}}|\widetilde{y}(f)|^2df,$$
and therefore,
$$\lambda_{N,\mathbb{W}}^{(\iota_+)}\leq\frac{\int_{\mathbb{W}}|\widetilde{y}(f)|^2df}{||y||_2^2}\leq\frac{1}{2}.$$

\subsection{Lower bound}
In the other direction, consider the minimax representation
$$\lambda_{N,\mathbb{W}}^{(l)} = \max_{S_{l+1}}~\min_{y\in S([N]),y\in S_{l+1}}\frac{\int_{\mathbb{W}}|\widetilde{y}(f)|^2df}{||y||_2^2}.$$
In order to find a lower bound for the eigenvalues, it suffices to pick an appropriate $(l+1)$-dimensional subspace $S_{l+1}\subset\ell_2(\mathbb{Z})$ and then find a uniform lower bound for the quantity above for all time-limited sequences $y\in S([N])$ inside $S_{l+1}$.
With $g$ as defined in (\ref{def_subspacefuc_g}), let the time-limited sequence $h\in\ell_2([N])$ be such that $h[n] = 1/\overline{g}[n]$ for all $n\in[N]$. We set
  $$L_- := \{n'\in\mathbb{Z}:-\lfloor\frac{N}{2}\rfloor\leq n'\leq\lfloor\frac{N-1}{2}\rfloor,~(\frac{n'}{N}-\frac{1}{2N},\frac{n'}{N}+\frac{1}{2N})\subset\mathbb{W}\},$$
and hence $\iota_- = \#L_-$. Let $S_{\iota_-}$ be the $\iota_-$-dimensional subspace of $\ell_2(\mathbb{Z})$ spanned by the functions $\mathcal{E}_{\frac{n'}{N}}h, n'\in L_-$, that is,
$$S_{\iota_-}:= \text{span}\left(\{\mathcal{E}_{\frac{n'}{N}}h\}_{ n'\in L_-}\right).$$
Suppose $y\in S_{\iota_-}$ (and hence $y\in\ell_2([N])$). Then we may write
$$y=\sum_{n'\in L_-}b_{n'}\mathcal{E}_{\frac{n'}{N}}h$$
for some coefficients $b_{n'}$.
Moreover,
$$y\odot\overline{g} = \sum_{n'\in L_-}b_{n'}{\bm e}_{\frac{n'}{N}}.$$
Noting that $\{\frac{1}{\sqrt{N}}{\bm e}_{\frac{n'}{N}},-\lfloor\frac{N}{2}\rfloor\leq n'\leq\lfloor\frac{N-1}{2}\rfloor\}$ forms an orthobasis for $\mathbb{C}^N$, we obtain
\e\sum_{n'\in L_-}|b_{n'}|^2 = N||y\odot\overline{g}||_2^2 = N\sum_{n=0}^{N-1}|y[n]\odot\overline{g}[n]|^2\geq\frac{1}{2}\sum_{n=0}^{N-1}|y[n]|^2=\frac{1}{2}||y||_2^2\label{eq_lowerbound_energy_g_2}
\nonumber\ee
since by definition, $g[n]\geq \frac{1}{\sqrt{2N}}$ for all $n\in[N]$. On the other hand,
$$b_{n'} = \sum_{n=0}^{N-1}\overline{g}[n]y[n]e^{-j2\pi \frac{n'}{N}n} = \langle y, \mathcal{E}_{\frac{n'}{N}} g\rangle.$$
Now using the same procedure  as in (\ref{eq_upperbound_energy_g_1}), one has
\e\begin{split}\sum_{n'\in L_-}|b_{n'}|^2
&=\sum_{n'\in L_-}|\langle y, \mathcal{E}_{\frac{n'}{N}} g\rangle|^2 \\
&=\sum_{n'\in L_-}|\int_{\frac{n'-1/2}{N}}^{\frac{n'+1/2}{N}}\widetilde{y}(f)\widetilde{\overline{g}}(\frac{n'}{N}-f)df|^2\\
&\leq\sum_{n'\in L_-}\left(||g||_2^2\int_{\frac{n'-1/2}{N}}^{\frac{n'+1/2}{N}}|\widetilde{y}(f)|^2df\right)\\
&\leq \int_{f\in\mathbb{W}}|\widetilde{y}(f)|^2df,
\end{split}\label{eq_upperbound_energy_g_2}
\nonumber\ee
where the last line holds since by construction, the set $\cup_{n'\in L_i}[\frac{n'}{N}-\frac{1}{2N},\frac{n'}{N}+\frac{1}{2N}]$ is a subset of the intervals $\mathbb{W}$. Altogether, we then conclude that for any $y\in S_{\iota_-}$ (and hence $y\in S([N])$),
$$\frac{1}{2}||y||_2^2\leq\int_{f\in\mathbb{W}}|\widetilde{y}(f)|^2df.$$
And hence
$$\lambda_{N,\mathbb{W}}^{(\iota_--1)}\geq \frac{\int_{f\in\mathbb{W}}|\widetilde{y}(f)|^2df}{||y||_2^2}\geq\frac{1}{2}. ~~~\square$$

\section{Proof of Theorem \ref{thm:EigConcentrationMulitband}}\label{proof:EigConcentrationMulitband}

\subsection{Proof of eigenvalues that cluster near zero}
\noindent\textbf{Proof.} Since $\bm{B}_{N,\mathbb{W}}=\sum_{i=0}^{J-1}\bm{E}_{f_i}\bm{B}_{N,W_i}\bm{E}_{f_i}^H$ , according to \cite{Horn1985MatrixAnalysis} (see pp. 181), the following holds
\begin{equation}
\lambda_{N,\mathbb{W}}^{(l)} \leq \sum_{i=0}^{J-1}
\lambda_{N,W_i}^{(l_i)}
\nonumber\end{equation}
for all $l_i\in[N], i \in[J]$ and $l=\sum_{i=0}^{J-1} l_i\in[N]$.

Fix $\epsilon\in(0,\frac{1}{|\mathbb{W}|}-1)$. For each $i\in[J]$, let $N_{1}(W_i,\epsilon)$, $C_3(W_i,\epsilon)$ and $C_4(W_i,\epsilon)$ be the constants  specified in Lemma \ref{lem:EigClusteringDPSS} with respect to $W_i$ and $\epsilon$. If we let $\overline{N}_1(\mathbb{W},\epsilon) = \max{\{N_{1}(W_i,\epsilon), ~\forall~ i \in [J]\}}$, then we have
$$\lambda_{N,W_i}^{(l_i)}\leq C_{3}(W_i,\epsilon)e^{-C_{4}(W_i,\epsilon)N},~ \forall~ l_i\geq \lceil 2NW_i(1+\epsilon)\rceil, i\in[J]$$
for all $N \geq \overline{N}_1(\mathbb{W},\epsilon)$. Hence, by choosing $l_i\geq \lceil 2NW_i(1+\epsilon)\rceil, ~\forall~ i\in[J]$, we have
\begin{equation}
\lambda_{N,\mathbb{W}}^{(l)} \leq \sum_{i=0}^{J-1} C_{3}(W_i,\epsilon)e^{-C_{4}(W_i,\epsilon)N} \leq \overline{C}_{3}(\mathbb{W},\epsilon)e^{-\overline{C}_4(\mathbb{W},\epsilon)N},
\nonumber\end{equation}
for all $N \geq \overline{N}_1(\mathbb{W},\epsilon)$ and $l\geq \sum_i \lceil 2NW_i(1+\epsilon)\rceil$,where $\overline{C}_3(\mathbb{W},\epsilon) = J\max{\{C_{3}(W_i,\epsilon), ~\forall~ i \in[J]\}}$ and $\overline{C}_4(\mathbb{W},\epsilon) = \min{\{C_{4}(W_i,\epsilon), ~\forall~ i \in[J]\}}$.  ~$\square$

\subsection{$\varepsilon$-pseudo eigenvalue and eigenvectors}
\begin{defn}($\varepsilon$-pseudo eigenvalue and eigenvector \cite{Reichel1992Pseudo-eigenvalues}) Let $\bm{X}\in\mathbb{C}^{N\times N}$ be any matrix and $\bm{u}\in\mathbb{C}^N$ be any vector with unit $l_2$-norm. Given $\varepsilon>0$, the number $\lambda\in\mathbb{C}$ and vector $\bm{u}\in\mathbb{C}^N$ are an $\varepsilon$-pseudo eigenpair of $\bm{X}$ if the following condition is satisfied:
$$
||(\bm{X}-\lambda\bm{I})\bm{u}||_2^2\leq \varepsilon.
$$
\end{defn}

 \begin{lem}Suppose $\mathbb{W}$ is a fixed finite union of $J$ pairwise disjoint intervals as defined in (\ref{eq:MultiBand}). Fix $\epsilon\in(0,1)$. For each $i\in[J]$, let $N_{0}(W_i,\epsilon)$ be the constant  specified in Lemma \ref{lem:EigClusteringDPSS} with respect to $W_i$ and $\epsilon$ and let $\widetilde{N}_0(\mathbb{W},\epsilon) = \max{\{N_{0}(W_i,\epsilon), ~\forall~ i \in [J]\}}$. Then for all $l_i\leq2NW_i(1-\epsilon), i \in [J]$ and $ N>\widetilde{N}_0(\mathbb{W},\epsilon)$,
 ($\lambda_{N,W_i}^{(l_i)}$, $\bm{E}_{f_i}\bm{s}_{N,W_i}^{(l_i)}$) is an $\varepsilon$-pseudo eigenpair of $\mathcal{I}_N\mathcal{B}_{\mathbb{W}}\mathcal{I}^*_N$ with $\varepsilon\leq 2C_{1}(W_i,\epsilon)e^{-C_{2}(W_i,\epsilon)N}$,
  or in detail
\begin{equation}
\mathcal{I}_N(\mathcal{B}_{\mathbb{W}}(\mathcal{I}^*_N(\bm{E}_{f_i}\bm{s}_{N,W_i}^{(l_i)})))=\lambda_{N,W_i}^{(l_i)}\bm{E}_{f_i}\bm{s}_{N,W_i}^{(l_i)}+\bm{o}_i^{(l_i)},
\nonumber\end{equation}
where $\bm{o}_i^{(l_i)} = \mathcal{I}_N(\mathcal{B}_{\mathbb{W}\setminus[f_i-W_i,f_i+W_i]}(\mathcal{I}^*_N(\bm{E}_{f_i}\bm{s}_{N,W_i}^{(l_i)})))$ and $||\bm{o}_i^{(l_i)}||_2^2\leq 2C_{1}(W_i,\epsilon)e^{-C_{2}(W_i,\epsilon)N}$. Here $\mathbb{W}\setminus[f_i-W_i,f_i+W_i] = \bigcup\limits_{i'\neq i}[f_{i'}-W_{i'},f_{i'}+W_{i'}]$ means the set difference between $\mathbb{W}$ and $[f_i-W_i,f_i+W_i]$, and $C_{1}(W_i,\epsilon)$  and $C_{2}(W_i,\epsilon)$ are the constants specified in Lemma \ref{lem:EigClusteringDPSS} corresponding to $W_i$ and $\epsilon$ for all $i\in[J]$.
\label{lem:approx_eigvect}\end{lem}
\vspace{0.3cm}
\noindent\textbf{Proof} (of Lemma \ref{lem:approx_eigvect}). According to the definition of the operator $\mathcal{I}_N\mathcal{B}_{\mathbb{W}}\mathcal{I}^*_N$,
\begin{equation}
\begin{split}
&\left(\mathcal{I}_N(\mathcal{B}_{\mathbb{W}}(\mathcal{I}^*_N(\bm{E}_{f_i}\bm{s}_{N,W_i}^{(l_i)})))\right)[m]\\
=&\sum_{n=0}^{N-1}\sum_{i'=0}^{J-1}e^{j2\pi f_{i'}(m-n)}\frac{\sin(2\pi W_{i'}(m-n))}{\pi(m-n)}e^{j2\pi f_in}\bm{s}_{N,W_i}^{(l_i)}[n]\\
=&e^{j2\pi f_im}\lambda^{(l_i)}_{N,W_i}\bm{s}_{N,W_i}^{(l_i)}[m]+\sum_{n=0}^{N-1}\sum_{i'=0,i'\neq i}^{J-1} e^{j2\pi f_{i'}(m-n)}\frac{\sin(2\pi W_{i'}(m-n))}{\pi(m-n)}e^{j2\pi f_in}\bm{s}_{N,W_i}^{(l_i)}[n]\\
=&e^{j2\pi f_im}\lambda^{(l_i)}_{N,W_i}\bm{s}_{N,W_i}^{(l_i)}[m]+\mathcal{I}_N(\mathcal{B}_{\mathbb{W}\setminus[f_i-W_i,f_i+W_i]}(\mathcal{I}^*_N(\bm{E}_{f_i}\bm{s}_{N,W_i}^{(l_i)})))[m].
\end{split}
\nonumber\end{equation}

In what follows, we will bound the energy of $\bm{o}_i^{(l_i)} = \mathcal{I}_N(\mathcal{B}_{\mathbb{W}\setminus[f_i-W_i,f_i+W_i]}(\mathcal{I}^*_N(\bm{E}_{f_i}\bm{s}_{N,W_i}^{(l_i)})))$ as
\begin{equation}
\begin{split}
||\bm{o}_i^{(l_i)}||_2^2 &= ||\mathcal{I}_N(\mathcal{B}_{\mathbb{W}\setminus[f_i-W_i,f_i+W_i]}(\mathcal{I}^*_N(\bm{E}_{f_i}\bm{s}_{N,W_i}^{(l_i)})))||_2^2\\
&\leq ||\mathcal{B}_{\mathbb{W}\setminus[f_i-W_i,f_i+W_i]}(\mathcal{I}^*_N(\bm{E}_{f_i}\bm{s}_{N,W_i}^{(l_i)}))||_2^2\\
&\leq ||\mathcal{B}_{[-\frac{1}{2},\frac{1}{2}]\setminus[f_i-W_i,f_i+W_i]}(\mathcal{I}^*_N(\bm{E}_{f_i}\bm{s}_{N,W_i}^{(l_i)}))||_2^2\\
&= ||\bm{s}_{N,W_i}^{(l_i)}||_2^2-||\mathcal{B}_{[f_i-W_i,f_i+W_i]}(\mathcal{I}^*_N(\bm{E}_{f_i}\bm{s}_{N,W_i}^{(l_i)}))||_2^2\\
&\leq
||\bm{s}_{N,W_i}^{(l_i)}||_2^2-||\mathcal{I}_N(\mathcal{B}_{[f_i-W_i,f_i+W_i]}(\mathcal{I}^*_N(\bm{E}_{f_i}\bm{s}_{N,W_i}^{(l_i)})))||_2^2\\
&\leq 1-(\lambda^{(l_i)}_{N,W_i})^2\leq 1-(1-C_{1}(W_i,\epsilon)e^{-C_{2}(W_i,\epsilon)N})^2\\
&=2C_{1}(W_i,\epsilon)e^{-C_{2}(W_i,\epsilon)N}-(C_1(W_i,\epsilon)e^{-C_2(W_i,\epsilon)N})^2\leq2C_{1}(W_i,\epsilon)e^{-C_{2}(W_i,\epsilon)N}
\end{split}\nonumber\end{equation} for all $l_i\leq \lfloor 2NW_i(1-\epsilon)\rfloor, i\in[J]$ and $N\geq \widetilde{N}_0(\mathbb{W},\epsilon)$. Here the second inequality in the sixth line follows simply from Lemma~\ref{lem:EigClusteringDPSS} since $\widetilde{N}_0(\mathbb{W},\epsilon)\geq N_0(W_i,\epsilon)$.~~ $\square$

\vspace{0.3cm}
Using this result, we now show the first $\approx N|\mathbb{W}|$ eigenvalues of $\mathcal{I}_N\mathcal{B}_{\mathbb{W}}\mathcal{I}^*_N$ are close to $1$.

\subsection{Proof of eigenvalues that cluster near one}

The main idea is to guarantee that the sum of the first $\approx N\left|\mathbb{W}\right|$ eigenvalues is sufficiently close $N|\mathbb{W}|$. Then we conclude that the first $\approx N|\mathbb{W}|$ eigenvalues cluster near one by applying the fact that the eigenvalues are upper bounded by $1$. First we state the following useful results.
\begin{lem}(\cite{DavenportWakin2012CSDPSS} Lemma 5.1)
Fix $\epsilon\in(0,1)$. Let $k_i = \lfloor 2NW_i(1-\epsilon)\rfloor, ~\forall~i
\in[J]$, and let $\bm{\Psi}$ be the dictionary as defined in (\ref{def_mul_dpss_dic}). Then for any pair of distinct columns $\bm{\psi}_1$ and $\bm{\psi}_2$ in $\bm{\Psi}$, we have
\e
\left|\langle \bm{\psi}_1, \bm{\psi}_2\rangle\right| \leq 3\sqrt{\widetilde{C}_1(\mathbb{W},\epsilon)}e^{-\frac{\widetilde{C}_2(\mathbb{W},\epsilon)}{2}N}
\label{eq:boundDPSSinnerProduct}\ee
and
$$\left\|\bm{\Psi}^H\bm{\Psi}\right\|_2 \leq 1+3N \sqrt{\widetilde{C}_1(\mathbb{W},\epsilon)}e^{-\frac{\widetilde{C}_2(\mathbb{W},\epsilon)N}{2}}$$
if $N\geq \widetilde{N}_0(\mathbb{W},\epsilon)$, where $\widetilde{C}_1(\mathbb{W},\epsilon) = \max{\{C_{1}(W_i,\epsilon), ~\forall~ i \in[J]\}}$ and $\widetilde{C}_2(\mathbb{W},\epsilon) = \min{\{C_{2}(W_i,\epsilon),~\forall~i \in [J]\}}$. Here $||\bm{\Psi}^H\bm{\Psi}||_2$ is the spectral norm (or largest singular value) of $\bm{\Psi}^H\bm{\Psi}$.
\label{lem_near_ortho_mul_dpss}\end{lem}

\begin{lem}(\cite{Horn1985MatrixAnalysis}) Let $\bm{X}\in\mathbb{C}^{N\times N}$ be a Hermitian matrix, and let $\lambda_0(\bm{X}),\lambda_1(\bm{X}),\ldots,\lambda_{N-1}(\bm{X})$ be its eigenvalues arranged in decreasing order. Then,
$$
\lambda_0(\bm{X})+\lambda_1(\bm{X})+\ldots+\lambda_{r-1}(\bm{X}) = \max_{\bm{U}\in\mathbb{C}^{N\times r},\bm{U}^H\bm{U}=\bm{I}_r} \textup{trace}(\bm{U}^H\bm{X}\bm{U}),
$$
where $\bm{I}_r$ is the $r\times r$ identity matrix with $1\leq r\leq N$.
\label{lem:hermitian_trace_equ}\end{lem}

Based on this result, we propose the following generalized result concerning the sum of the first $r$ eigenvalues.
\begin{lem}Let $\bm{X}\in\mathbb{C}^{N\times N}$ be a positive-semidefinite (PSD) matrix, and let $\lambda_0(\bm{X}),\lambda_1(\bm{X}),\ldots,\lambda_{N-1}(\bm{X})$ be its eigenvalues arranged in decreasing order. Then, for any matrix $\bm{M}\in\bm{C}^{N\times r}, 1\leq r\leq N$, the following inequality holds
$$
\lambda_0(\bm{X})+\lambda_1(\bm{X})+\ldots+\lambda_{r-1}(\bm{X}) \geq \textup{trace}(\bm{M}^H\bm{X}\bm{M})/\|\bm{M}^H\bm{M}\|_2.
$$
\label{lem:hermitian_trace_inequ}\end{lem}
\noindent\textbf{Proof} (of Lemma \ref{lem:hermitian_trace_inequ}). Let $\sigma_0(\bm{M}),\ldots,\sigma_{r-1}(\bm{M})$ denote the decreasing singular values of the matrix $\bm{M}$.
Denote $\bm{M} = \bm{U}_r\bm{\Sigma}_r\bm{V}^H_r$ as the truncated SVD of $\bm{M}$, where $\bm{\Sigma}_r$ is an $r\times r$ diagonal matrix with $\sigma_0(\bm{M}),\ldots,\sigma_{r-1}(\bm{M})$ along its diagonal.

Now applying Lemma \ref{lem:hermitian_trace_equ}, we obtain
\e\begin{split}
\sum_{l = 0}^{r-1} \lambda_l(\bm{X}) &\geq \text{trace}(\bm{U}_r^H\bm{X}\bm{U}_r)\\
& \geq \text{trace}(\bm{\Sigma}_r\bm{U}_r^H\bm{X}\bm{U}_r\bm{\Sigma}_r)/(\sigma_0(\bm{M}))^2\\
&=\text{trace}(\bm{V_r}\bm{\Sigma_r}\bm{U}_r^H\bm{X}\bm{U}_r\bm{\Sigma}_r\bm{V}^H_r)/\|\bm{M}^H\bm{M}\|_2\\
&=\text{trace}(\bm{M}^H\bm{X}\bm{M})/\|\bm{M}^H\bm{M}\|_2,
\end{split}\nonumber\ee
where the first line follows directly from Lemma \ref{lem:hermitian_trace_equ}, the second line is obtained because $\bm{U}_r^H\bm{X}\bm{U}_r$ is PSD and hence its main diagonal elements are non-negative, and the third line follows because $\bm{V}_r$ is an orthobasis and $(\sigma_0(\bm{M}))^2 = \|\bm{M}^H\bm{M}\|_2$. ~ $\square$

We are now ready to prove the main part. Fix $\epsilon\in(0,1)$. Let $k_i = \lfloor 2NW_i(1-\epsilon)\rfloor, \forall i\in[J]$, and let $\bm{\Psi}$ be the dictionary as defined in (\ref{def_mul_dpss_dic}). We have
\e\begin{split}
\sum_{l=0}^{J-1+\sum_i \lfloor 2NW_i(1-\epsilon)\rfloor}\lambda_{N,\mathbb{W}}^{(l)} & \geq \text{trace}\left(\bm{\Psi}^H\bm{B}_{N,\mathbb{W}}\bm{\Psi}\right)/\left\|\bm{\Psi}^H\bm{\Psi}\right\|_2\\
& = \left(\sum_{i=0}^{J-1}\sum_{l_i= 0}^{\lfloor 2NW_i(1-\epsilon)\rfloor}\left((\bm{E}_{f_i}\bm{s}_{N,W_i}^{(l_i)})^H\mathcal{I}_N(\mathcal{B}_{\mathbb{W}}(\mathcal{I}^*_N(\bm{E}_{f_i}\bm{s}_{N,W_i}^{(l_i)})))\right)\right)/\left\|\bm{\Psi}^H\bm{\Psi}\right\|_2\\
& = \left(\sum_{i=0}^{J-1}\sum_{l_i= 0}^{\lfloor 2NW_i(1-\epsilon)\rfloor}\left(\left(\bm{E}_{f_i}\bm{s}_{N,W_i}^{(l_i)}\right)^H\left(\lambda_{N,W_i}^{(l_i)}\bm{E}_{f_i}\bm{s}_{N,W_i}^{(l_i)}+\bm{o}_i^{(l_i)}\right)\right)\right)/\left\|\bm{\Psi}^H\bm{\Psi}\right\|_2\\
& \geq \left(\sum_{i=0}^{J-1}\sum_{l_i= 0}^{\lfloor 2NW_i(1-\epsilon)\rfloor}\left(\lambda_{N,W_i}^{(l_i)}-\|\bm{o}_{i}^{(l_i)}\|_2\right)\right)/\left\|\bm{\Psi}^H\bm{\Psi}\right\|_2\\
& \geq \frac{\left(\sum_{i=0}^{J-1}\sum_{l_i= 0}^{\lfloor 2NW_i(1-\epsilon)\rfloor}\left(1-C_1(W_i,\epsilon)e^{-C_2(W_i,\epsilon)N}-\sqrt{2}\sqrt{C_1(W_i,\epsilon)}e^{-\frac{C_2(W_i,\epsilon)}{2}N}\right)\right)}{\left(1+3N \sqrt{\widetilde{C}_1(\mathbb{W},\epsilon)}e^{-\frac{\widetilde{C}_2(\mathbb{W},\epsilon)N}{2}}\right)}\\
& \geq \frac{\left(\sum_{i=0}^{J-1}\sum_{l_i= 0}^{\lfloor 2NW_i(1-\epsilon)\rfloor}\left(1-\widetilde{C}_1(\mathbb{W},\epsilon)e^{-\widetilde{C}_2(\mathbb{W},\epsilon)N}-\sqrt{2}\sqrt{\widetilde{C}_1(\mathbb{W},\epsilon)}e^{-\frac{\widetilde{C}_2(\mathbb{W},\epsilon)}{2}N}\right)\right)}{\left(1+3N \sqrt{\widetilde{C}_1(\mathbb{W},\epsilon)}e^{-\frac{\widetilde{C}_2(\mathbb{W},\epsilon)N}{2}}\right)}\\
& \geq \frac{J+\sum_i \lfloor 2NW_i(1-\epsilon)\rfloor-3NC_5(\mathbb{W},\epsilon)e^{-\frac{\widetilde{C}_2(\mathbb{W},\epsilon)}{2}N}}{1+3N C_5(\mathbb{W},\epsilon)e^{-\frac{\widetilde{C}_2(\mathbb{W},\epsilon)N}{2}}}\\
&=\frac{\left(J+\sum_i \lfloor 2NW_i(1-\epsilon)\rfloor-3NC_5(\mathbb{W},\epsilon)e^{-\frac{\widetilde{C}_2(\mathbb{W},\epsilon)}{2}N}\right)\left(1-3N C_5(\mathbb{W},\epsilon)e^{-\frac{\widetilde{C}_2(\mathbb{W},\epsilon)N}{2}}\right)}{\left(1+3N C_5(\mathbb{W},\epsilon)e^{-\frac{\widetilde{C}_2(\mathbb{W},\epsilon)N}{2}}\right)\left(1-3N C_5(\mathbb{W},\epsilon)e^{-\frac{\widetilde{C}_2(\mathbb{W},\epsilon)N}{2}}\right)}\\&
\geq \frac{J+\sum_i \lfloor 2NW_i(1-\epsilon)\rfloor-6N^2C_5(\mathbb{W},\epsilon)e^{-\frac{\widetilde{C}_2(\mathbb{W},\epsilon)}{2}N}
+\left(3NC_5(\mathbb{W},\epsilon)e^{-\frac{\widetilde{C}_2(\mathbb{W},\epsilon)}{2}N}\right)^2}
{1-\left(3N C_5(\mathbb{W},\epsilon)e^{-\frac{\widetilde{C}_2(\mathbb{W},\epsilon)N}{2}}\right)^2}\\&
\geq J+\sum_i \lfloor 2NW_i(1-\epsilon)\rfloor - 6N^2C_5(\mathbb{W},\epsilon)e^{-\frac{\widetilde{C}_2(\mathbb{W},\epsilon)}{2}N}
\end{split}\nonumber\ee
for all $N\geq \max\{\widetilde{N}_0(\mathbb{W},\epsilon),N'(\mathbb{W},\epsilon)\}$, where $N'(\mathbb{W},\epsilon) =\max\{(\frac{4}{C_2(\mathbb{W},\epsilon)})^2,~\frac{4}{C_2(\mathbb{W},\epsilon)}\log(3C_5(\mathbb{W},\epsilon))\}$ is the constant such that $3N C_5(\mathbb{W},\epsilon)e^{-\frac{\widetilde{C}_2(\mathbb{W},\epsilon)N}{2}}<1$ for all $N\geq N'(\mathbb{W},\epsilon)$.\footnote{This can be verified as $3N C_5(\mathbb{W},\epsilon)e^{-\frac{\widetilde{C}_2(\mathbb{W},\epsilon)N}{2}} = 3 C_5(\mathbb{W},\epsilon)e^{-N(\frac{\widetilde{C}_2(\mathbb{W},\epsilon)}{2}-\frac{\log N}{N})}\leq 3 C_5(\mathbb{W},\epsilon)e^{-N\frac{\widetilde{C}_2\mathbb{W},\epsilon)}{4}}\leq 1$ for all $N\geq \max\{(\frac{4}{C_2(\mathbb{W},\epsilon)})^2,~\frac{4}{C_2(\mathbb{W},\epsilon)}\log(3C_5(\mathbb{W},\epsilon))\}$. Here the first inequality follows because $\frac{\log N}{N}\leq \frac{1}{N^{1/2}}\leq \frac{C_2(\mathbb{W},\epsilon)}{4}$ for all $N\geq (\frac{4}{C_2(\mathbb{W},\epsilon)})^2$. } Here the first line follows directly from Lemma \ref{lem:hermitian_trace_inequ}, the second line follows because $\text{trace}\left(\bm{\Psi}^H\bm{B}_{N,\mathbb{W}}\bm{\Psi}\right) = \text{trace}\left(\sum_{i=0}^{J-1}\bm{\Psi}_i^H\bm{B}_{N,\mathbb{W}}\bm{\Psi}_i\right)$ and $\bm{B}_{N,\mathbb{W}}$ is equivalent to $\mathcal{I}_N\mathcal{B}_{\mathbb{W}}\mathcal{I}^*_N$, the third line follows from Lemma \ref{lem:approx_eigvect}, the fourth line follows from the Cauchy-Schwarz inequality which indicates that $|(\bm{E}_{f_i}\bm{s}_{N,W_i}^{(l_i)})^H\bm{o}_{i}^{(l_i)}|\leq ||\bm{E}_{f_i}\bm{s}_{N,W_i}^{(l_i)}||_2||\bm{o}_{i}^{(l_i)}||_2 = ||\bm{o}_{i}^{(l_i)}||_2$, the fifth line follows from Lemmas \ref{lem:EigClusteringDPSS}, \ref{lem:approx_eigvect} and \ref{lem_near_ortho_mul_dpss}, the seventh line follows by setting $C_5(\mathbb{W},\epsilon) = \max\{\widetilde{C}_1(\mathbb{W},\epsilon),\sqrt{\widetilde{C}_1(\mathbb{W},\epsilon)}\}$, the ninth line follows because $J+\sum_i \lfloor 2NW(1-\epsilon)\rfloor\leq N$, and the last line follows because by assumption $3N C_5(\mathbb{W},\epsilon)e^{-\frac{\widetilde{C}_2(\mathbb{W},\epsilon)N}{2}}<1$.

By noting that $0<\lambda_{N,\mathbb{W}}^{(N-1)}\leq\lambda_{N,\mathbb{W}}^{(0)}<1$ from Lemma \ref{lem:PSD}, we acquire
\e\begin{split}
\lambda_{N,\mathbb{W}}^{(l)} &
 =\left(\sum_{l'=0}^{J-1+\sum_i \lfloor 2NW_i(1-\epsilon)\rfloor}\lambda_{N,\mathbb{W}}^{(l')}\right) - \left(\sum_{l'=0,l'\neq l}^{J-1+\sum_i \lfloor 2NW_i(1-\epsilon)\rfloor}\lambda_{N,\mathbb{W}}^{(l')}\right)\\
&\geq \left(\sum_{l'=0}^{J-1+\sum_i \lfloor 2NW_i(1-\epsilon)\rfloor}\lambda_{N,\mathbb{W}}^{(l')}\right) - \left(J-1+\sum_i\lfloor 2NW_i(1-\epsilon)\rfloor\right)\\
&\geq 1-6N^2C_5(\mathbb{W},\epsilon)e^{-\frac{\widetilde{C}_2(\mathbb{W},\epsilon)}{2}N}
\end{split}\nonumber\ee
for all $l\leq J-1+\sum_i \lfloor 2NW_i(1-\epsilon)\rfloor$, where the second line follows by setting $\lambda_{N,\mathbb{W}}^{(l')},~l'\neq l$ to $1$. Fix $\mathbb{W}$ and $\epsilon$. It is always possible to find a constant $N'$ such that $3N C_5(\mathbb{W},\epsilon)e^{-\frac{\widetilde{C}_2(\mathbb{W},\epsilon)N}{2}}<1$ for all $N\geq N'$. Now, for convenience, we set $\overline{C}_1(\mathbb{W},\epsilon) = 6C_5(\mathbb{W},\epsilon)$, $\overline{C}_2(\mathbb{W},\epsilon) = \frac{\widetilde{C}_2(\mathbb{W},\epsilon)}{2}$, and $\overline{N}_0(\mathbb{W},\epsilon) = \max\{\widetilde{N}_0(\mathbb{W},\epsilon),N'\}$.
This completes the proof of Theorem~\ref{thm:EigConcentrationMulitband}. ~~ $\square$

\section{Proof of Theorem \ref{thm_subspace_angle_1}}\label{proof_thm_subspace_angle_1}

\textbf{Proof.}
First denote the eigen-decomposition of $\bm{B}_{N,\mathbb{W}}$ as
$$
\bm{B}_{N,\mathbb{W}} = \bm{U}_{N,\mathbb{W}}\bm{\Lambda}_{N,\mathbb{W}}\bm{U}_{N,\mathbb{W}}^H,
$$
where $\bm{\Lambda}_{N,\mathbb{W}}$ is an $N\times N$ diagonal matrix whose diagonal elements are the eigenvalues $\lambda_{N,\mathbb{W}}^{(0)},\lambda_{N,\mathbb{W}}^{(1)},\ldots,\lambda_{N,\mathbb{W}}^{(N-1)}$ and $\bm{U}_{N,\mathbb{W}}$ is a square ($N\times N$) matrix defined by
$$
\bm{U}_{N,\mathbb{W}} :=[\bm{u}_{N,\mathbb{W}}^{(0)} ~ \bm{u}_{N,\mathbb{W}}^{(1)} ~ \ldots ~ \bm{u}_{N,\mathbb{W}}^{(N-1)}].
$$
Also let $\bm{a} = \bm{U}_{N,\mathbb{W}}^H\bm{\psi}$ be the coefficients of $\bm{\psi}$ represented by $\bm{U}_{N,\mathbb{W}}$.

Fix $\epsilon\in(0,\min\{1,\frac{1}{|\mathbb{W}|}-1\})$. Suppose $\bm\psi$ is a column of $\bm{\Psi}_i$ for some particular $i\in[J]$.
Now from Lemma \ref{lem:approx_eigvect}, we have
\e
\bm{B}_{N,\mathbb{W}}\bm{\psi}=\lambda_{N,W_i}^{(l_i)}\bm{\psi}+\bm{o}_i^{(l_i)}
\nonumber\ee
for some $l_i\leq \lfloor 2NW_i(1-\epsilon)\rfloor$.

Plugging the eigen-decomposition of the matrix $\bm{U}_{N,\mathbb{W}}$ into the above equation, we require
\e
\bm{\Lambda}_{N,\mathbb{W}}\bm{a} =  \lambda_{N,W_i}^{(l_i)}\bm{a}+\widehat{\bm{o}}_i^{(l_i)},
\nonumber\ee
where $\widehat{\bm{o}}_i^{(l_i)} = \bm{U}_{N,\mathbb{W}}^H\bm{o}_i^{(l_i)}$.
The elementary form of the above equation is
$$\lambda_{N,\mathbb{W}}^{(m)}\bm{a}[m] = \lambda_{N,W_i}^{(l_i)}\bm{a}[m]+ \widehat{\bm{o}}_i^{(l_i)}[m]$$ for all $m\in[N]$.

Now we have
\e\begin{split}||\bm\psi-\bm{P}_{{\bm{\Phi}}}\bm\psi||_2^2 & = \sum_{m=\sum_i \lceil2NW_i(1+\epsilon)\rceil}^{N-1}|\bm{a}[m]|^2 = \sum_{m=\sum_i \lceil2NW_i(1+\epsilon)\rceil}^{N-1}\frac{\left|\widehat{\bm{o}}_i^{(l_i)}[m]\right|^2}{\left|\lambda_{N,W_i}^{(l_i)}-\lambda_{N,\mathbb{W}}^{(m)}\right|^2}\\
& \leq \frac{\sum_{m=\sum_i \lceil2NW_i(1+\epsilon)\rceil}^{N-1}\left|\widehat{\bm{o}}_i^{(l_i)}[m]\right|^2}{\left(1-\widetilde{C}_1(\mathbb{W},\epsilon)e^{-\widetilde{C}_2(\mathbb{W},\epsilon)N}-\overline{C}_3(\mathbb{W},\epsilon)e^{-\overline{C}_4(\mathbb{W},\epsilon)N}\right)^2}\\
& \leq \frac{||\bm{o}_i^{(l_i)}||^2}{\left(1-\widetilde{C}_1(\mathbb{W},\epsilon)e^{-\widetilde{C}_2(\mathbb{W},\epsilon)N}-\overline{C}_3(\mathbb{W},\epsilon)e^{-\overline{C}_4(\mathbb{W},\epsilon)N}\right)^2}\\
& \leq \frac{2\widetilde{C}_1(\mathbb{W},\epsilon)e^{-\widetilde{C}_2(\mathbb{W},\epsilon)N}}{\left(1-\widetilde{C}_1(\mathbb{W},\epsilon)e^{-\widetilde{C}_2(\mathbb{W},\epsilon)N}-\overline{C}_3(\mathbb{W},\epsilon)e^{-\overline{C}_4(\mathbb{W},\epsilon)N}\right)^2}\\
\end{split}\label{eq:boundProjpsi}\ee
for all $N\geq \max\{\overline{N}_0(\mathbb{W},\epsilon),\overline{N}_1(\mathbb{W},\epsilon)\}$, where the second line follows by bounding the $\lambda_{N,W_i}^{(l_i)}$ term using $1-C_1(W_i,\epsilon)e^{-C_2(W_i,\epsilon)N}$ (which is not less than $1-\widetilde{C}_1(\mathbb{W},\epsilon)e^{-\widetilde{C}_2(\mathbb{W},\epsilon)N}$) from Lemma~\ref{lem:EigClusteringDPSS} and bounding the $\lambda_{N,\mathbb{W}}^{(m)}$ terms using Theorem \ref{thm:EigConcentrationMulitband}, and the fourth line follows because $||\bm{o}_i^{(l_i)}||^2\leq 2C_1(W_i,\epsilon)e^{-C_2(W_i,\epsilon)N}\leq 2\widetilde{C}_1(\mathbb{W},\epsilon)e^{-\widetilde{C}_2(\mathbb{W},\epsilon)N}$.

The following general result will help in extending~\eqref{eq:boundProjpsi} to an angle between the subspaces.
\begin{lem}
Let $\mathcal{S}_{\bm{U}}$ and $\mathcal{S}_{\bm{V}}$ be the subspaces spanned by the columns of the matrices $\bm{U}\in \mathbb{C}^{N\times q}$ and $\bm{V}\in\mathbb{C}^{N\times r}$, respectively. Here $r \leq q \leq N$. Suppose each column of $\bm V$ is normalized so that $\|\bm{v}_l\|_2 = 1$ and is close to $\mathcal{S}_{\bm{U}}$ such that for some $\delta_1$,  $\|\bm{v}_l - \bm P_{\bm U}\bm{v}_l\|_2^2\leq \delta_1$ for all $l\in [r]$. Furthermore, suppose the columns of $\bm V$ are approximately orthogonal to each other such that for some $\delta_2$,  $\left|\langle\bm{v}_k, \bm{v}_l\rangle\right|\leq \delta_2$ for all $k\neq l$. Then we have
$$\cos(\Theta_{\mathcal{S}_{\bm U}\mathcal{S}_{\bm V}})\geq\sqrt{\frac{1 - \delta_1- N\left(\delta_2+\sqrt{\delta_1}\right)}{1 + N \delta_2}}.$$
\label{lem:BoundSubspaceAngle}\end{lem}
\noindent\textbf{Proof} (of Lemma \ref{lem:BoundSubspaceAngle}). Any $\bm v\in \mathcal{S}_{\bm{V}}$ can be written as a linear combination of $\bm v_l$ in the form $\bm v = \sum_l \alpha_l \bm v_l$. We first bound the $l_2$ norm of $\bm v$ by
\e\begin{split}
\|\bm v\|_2^2 & = \|\sum_{l=0}^{r-1} \alpha_l \bm v_l\|_2^2\\
& = \sum_{l=0}^{r-1} |\alpha_l|^2\|\bm v_l\|_2^2 + \sum_{l=0}^{r-1}\sum_{k=0, k\neq l}^{r-1}\langle \alpha_l\bm v_l,\alpha_k \bm v_k \rangle\\
&\leq \sum_{l=0}^{r-1} |\alpha_l|^2 + \sum_{l=0}^{r-1}\sum_{k=0, k\neq l}^{r-1}|\alpha_l||\alpha_k|\delta_2\\
& \leq \sum_{l=0}^{r-1} |\alpha_l|^2 + \sum_{l=0}^{r-1}\sum_{k=0, k\neq l}^{r-1}\frac{|\alpha_l|^2+|\alpha_k|^2}{2}\delta_2\\
& = \left(\sum_{l=0}^{r-1} |\alpha_l|^2\right)\left(1 + (r-1)\delta_2\right)\leq \left(\sum_{l=0}^{r-1} |\alpha_l|^2\right)\left(1 + N\delta_2\right),
\end{split}\nonumber\ee
where the third line follows from the hypothesis that $\left|\langle\bm{v}_k, \bm{v}_l\rangle\right|\leq \delta_2$ for all $k\neq l$. Similarly,
\e\begin{split}
\|\bm P_{\bm U}\bm{v}\|_2^2 & = \|\sum_{l=0}^{r-1} \bm P_{\bm U}\left(\alpha_l \bm v_l\right)\|_2^2\\
& = \sum_{l=0}^{r-1} |\alpha_l|^2\| \bm P_{\bm U}\bm v_l\|_2^2 + \sum_{l=0}^{r-1}\sum_{k=0, k\neq l}^{r-1}\left\langle \alpha_l\bm P_{\bm U}\bm v_l,\alpha_k \bm P_{\bm U}\bm v_k \right\rangle\\
& = \sum_{l=0}^{r-1} |\alpha_l|^2\| \bm P_{\bm U}\bm v_l\|_2^2 + \sum_{l=0}^{r-1}\sum_{k=0, k\neq l}^{r-1}\left\langle \alpha_l \bm v_l,\alpha_k \left(\bm v_k - (\bm v_k -\bm P_{\bm U}\bm v_k)\right) \right\rangle\\
&\geq \sum_{l=0}^{r-1} |\alpha_l|^2\left(1-\delta_1\right) - \sum_{l=0}^{r-1}\sum_{k=0, k\neq l}^{r-1}|\alpha_l||\alpha_k|\left(\delta_2+\sqrt{\delta_1}\right)\\
& = \left(\sum_{l=0}^{r-1} |\alpha_l|^2\right)\left(1 - \delta_1- (r-1)\left(\delta_2+\sqrt{\delta_1}\right)\right)\geq \left(\sum_{l=0}^{r-1} |\alpha_l|^2\right)\left(1 - \delta_1- N\left(\delta_2+\sqrt{\delta_1}\right)\right),
\end{split}\nonumber\ee
where the fourth line follows because  $\langle \bm v_l, \bm v_k -\bm P_{\bm U}\bm v_k\rangle\leq \|\bm v_l\|_2\|\bm v_k - \bm P_{\bm U}\bm v_k\|_2\leq \sqrt{\delta_1}$ and $\left|\langle\bm{v}_k, \bm{v}_l\rangle\right|\leq \delta_2$ for all $k\neq l$.

Therefore, for any non-zero vector $\bm v\in \mathcal{S}_{\bm{V}}$ we have
$$\frac{\|\bm P_{\bm U}\bm{v}\|_2^2}{\|\bm v\|_2^2}\geq \frac{1 - \delta_1- N\left(\delta_2+\sqrt{\delta_1}\right)}{1 + N \delta_2}.~~~\square$$

Finally, (\ref{eq_subsapce_angle_1_2}) follows from Lemma~\ref{lem:BoundSubspaceAngle} by replacing $\bm U$ with $\bm \Phi$ and $\bm V$ with $\bm \Psi$, and assigning $\delta_1$ with the upper bound in (\ref{eq:boundProjpsi}) and  $\delta_2$ with the upper bound in (\ref{eq:boundDPSSinnerProduct}). ~~ $\square$

\section{Proof of Theorem \ref{thm_subspace_approx_2}} \label{proof_thm_subspace_approx_2}

\noindent\textbf{Proof.}
For each $i\in[J]$, define $\overline{\bm{\Psi}}_i = [\bm{E}_{f_i}\bm{S}_{N,W_i}\sqrt{\bm{\Lambda}_{N,W_i}}]_{k_i}$ for some given $k_i\in\{1,2,\ldots,N\}$. We construct the scaled multiband modulated DPSS matrix $\overline{\bm{\Psi}}$ by\footnote{Hogan and Lakey~\cite{Hogan2014FramePropertiesPSWF} considered the scaled and shifted Prolate Spheroidal Wave Fuctions (PSWF's) and provided conditions on a shift parameter such that the scaled and shifted PSWF's form a frame or a Riesz basis for the Paley-Wiener space.}
\e
\overline{\bm{\Psi}}:=[\overline{\bm{\Psi}}_0 ~ \overline{\bm{\Psi}}_1 ~ \cdots ~ \overline{\bm{\Psi}}_{J-1}].
\label{def_scaled_mul_dpss_dic}\ee
The main idea is to bound $\left\|\bm{P}_{\bm{\Psi}} \bm{u}_{N,\mathbb{W}}^{(l)}\right\|_2$ using $\left\|\overline{\bm{\Psi}}~\overline{\bm{\Psi}}^H\bm{u}_{N,\mathbb{W}}^{(l)}\right\|_2$. In order to use this argument, we first give out some useful results.

\begin{lem}
Suppose $\overline{\bm{\Psi}}$ is the matrix defined in (\ref{def_scaled_mul_dpss_dic}) with some given $k_i\in\{1,2,\ldots,N\}, \forall i\in[J]$. Then
$$\left\|\overline{\bm{\Psi}}\right\|_2\leq 1.\label{eq_scaled_dic_bounded}$$
\label{lem_scaled_dic_bounded}\end{lem}
\textbf{Proof} (of Lemma \ref{lem_scaled_dic_bounded})
Let $\bm{y}\in \mathbb{C}^N$. Then
\e\begin{split}
\left\|\overline{\bm{\Psi}}^H\bm{y}\right\|_2^2 &=\sum_{i=0}^{J-1}\sum_{l_i=0}^{k_i-1}|\langle \bm{y},\bm{E}_{f_i}\sqrt{\lambda_{N,W_i}^{(l_i)}}\bm{s}_{N,W_i}^{(l_i)}\rangle|^2\\ &=\sum_{i=0}^{J-1}\sum_{l_i=0}^{k_i-1}\langle \bm{y},\bm{E}_{f_i}\sqrt{\lambda_{N,W_i}^{(l_i)}}\bm{s}_{N,W_i}^{(l_i)}\rangle\langle \bm{E}_{f_i}\sqrt{\lambda_{N,W_i}^{(l_i)}}\bm{s}_{N,W_i}^{(l_i)}, \bm{y}\rangle \\
&=\sum_{i=0}^{J-1}\sum_{l_i=0}^{k_i-1} \bm{y}^H\bm{E}_{f_i}\bm{s}_{N,W_i}^{(l_i)}\lambda_{N,W_i}^{(l_i)}(\bm{s}_{N,W_i}^{(l_i)})^H \bm{E}_{f_i}^H\bm{y}\\
&\leq\sum_{i=0}^{J-1}\sum_{l_i=0}^{N-1} \bm{y}^H\bm{E}_{f_i}\bm{s}_{N,W_i}^{(l_i)}\lambda_{N,W_i}^{(l_i)}(\bm{s}_{N,W_i}^{(l_i)})^H \bm{E}_{f_i}^H\bm{y}\\
&=\sum_{i=0}^{J-1} \bm{y}^H\bm{E}_{f_l}\mathcal{I}_N(\mathcal{B}_{W_i}(\mathcal{I}^*_N( \bm{E}_{f_i}^H\bm{y})))=\sum_{i=0}^{J-1} \langle\mathcal{I}_N(\mathcal{B}_{W_i}(\mathcal{I}^*_N( \bm{E}_{f_i}^H\bm{y}))),\bm{E}_{f_i}^H\bm{y}\rangle\\&=\sum_{i=0}^{J-1} \langle\mathcal{B}_{W_i}(\mathcal{I}^*_N( \bm{E}_{f_i}^H\bm{y})),\mathcal{I}^*_N(\bm{E}_{f_i}^H\bm{y})\rangle=\sum_{i=0}^{J-1} \langle\mathcal{B}_{W_i}(\mathcal{I}^*_N( \bm{E}_{f_i}^H\bm{y})),\mathcal{B}_{W_i}(\mathcal{I}^*_N(\bm{E}_{f_l}^H\bm{y}))\rangle=\sum_{i=0}^{J-1} ||\mathcal{B}_{W_i}(\mathcal{I}^*_N( \bm{E}_{f_i}^H\bm{y}))||_2^2\\
&=\sum_{i=0}^{J-1}\int_{f_i-W_i}^{f_i+W_i}|\widetilde{\bm{y}}(f)|^2df=\int_{-1/2}^{1/2}\left(\sum_{i=0}^{J-1}\mathbbm{1}_{[f_i-W_i,f_i+W_i)}(f)\right)|\widetilde{\bm{y}}(f)|^2df
\label{eq_proof_scaled_dic_bounded},\end{split}\nonumber\ee
where the fourth line follows because $\bm{y}^H\bm{E}_{f_i}\bm{s}_{N,W_i}^{(l_i)}\lambda_{N,W_i}^{(l_i)}(\bm{s}_{N,W_i}^{(l_i)})^H \bm{E}_{f_i}^H\bm{y} = ||\sqrt{\lambda_{N,W_i}^{(l_i)}}(\bm{s}_{N,W_i}^{(l_i)})^H \bm{E}_{f_i}^H\bm{y}||_2^2\geq 0$, the fifth line follows because $\sum_{l_i=0}^{N-1}\bm{s}_{N,W_i}^{(l_i)}\lambda_{N,W_i}^{(l_i)}(\bm{s}_{N,W_i}^{(l_i)})^H \bm{x} =\mathcal{I}_N(\mathcal{B}_{W_i}(\mathcal{I}^*_N(\bm{x})))$, and we use $\widetilde{\bm{y}}(f) = \sum_{n=0}^{N-1}\bm{y}[n]e^{-j2\pi fn}$ as the DTFT of $\mathcal{I}^*_N(\bm{y})$ in the last three equations.

Noting that $\sum_{i=0}^{J-1}\mathbbm{1}_{[f_i-W_i,W_i+f_i)}(f)\leq 1$ for all $f\in[-\frac{1}{2},\frac{1}{2}]$ since we assume there is no overlap between each interval $[f_i-W_i,W_i+f_i)$, we conclude
$$||\overline{\bm{\Psi}}^H\bm{y}||_2^2\leq \int_{-1/2}^{1/2}|\widetilde{\bm y}(f)|^2df = ||\bm{y}||_2^2$$
and
$$||\overline{\bm{\Psi}}||_2\leq 1. ~~\square$$

\begin{lem}
For any $k_i\in\{1,2,\ldots,N\}, i\in[J]$, let $\bm{\Psi}$ and $\overline{\bm{\Psi}}$ be the matrices defined in (\ref{def_mul_dpss_dic}) and (\ref{def_scaled_mul_dpss_dic}) respectively. Then for any $\bm{y}\in \mathbb{C}^{N\times 1}$,
\e||\bm{P}_{\bm{\Psi}}\bm{y}||_2\geq ||\overline{\bm{\Psi}}~\overline{\bm{\Psi}}^H\bm{y}||_2.
\label{eq_proj_multi_bounded}\ee
\label{lem_proj_multi_bounded}\end{lem}

\textbf{Proof} (of Lemma \ref{lem_proj_multi_bounded}) Let $\overline{\bm{\Psi}} = \bm{U}_{\overline{\bm{\Psi}}}\Sigma_{\overline{\bm{\Psi}}}\bm{V}_{\overline{\bm{\Psi}}}^H$ be a reduced SVD of $\overline{\bm{\Psi}}$, where both $\bm{U}_{\overline{\bm{\Psi}}}$ and $\bm{V}_{\overline{\bm{\Psi}}}$ are orthonormal matrices of the proper dimension, and $\bm{\Sigma}_{\overline{\bm{\Psi}}}$ is a diagonal matrix whose diagonal elements are the non-zero singular values of $\overline{\bm{\Psi}}$. We have
\e\begin{split}
||\overline{\bm{\Psi}}~\overline{\bm{\Psi}}^H\bm{y}||_2 &= ||\bm{U}_{\overline{\bm{\Psi}}}\Sigma_{\overline{\bm{\Psi}}}^2\bm{U}_{\overline{\bm{\Psi}}}^H\bm{y}||_2\\
& \leq ||\bm{U}_{\overline{\bm{\Psi}}}^H\bm{y}||_2\\
& = ||\bm{U}_{\overline{\bm{\Psi}}}\bm{U}_{\overline{\bm{\Psi}}}^H\bm{y}||_2\\
&= ||\bm{P}_{\bm{\Psi}}\bm{y}||_2
\end{split}\nonumber\ee
where the second lines follows because $||\overline{\bm{\Psi}}||_2\leq 1$ and hence the diagonal elements $\bm{\Sigma}_{\overline{\bm{\Psi}}}$ are bounded above by $1$, and the fourth line follows because each column in $\overline{\bm{\Psi}}$ is in also $\bm{\Psi}$ and hence $||\bm{P}_{\bm{\Psi}}\bm{y}||_2=||\bm{P}_{\bm{U}_{\overline{\Psi}}}\bm{y}||_2$. $\square$

\vspace{0.3cm}
Now we turn to prove Theorem \ref{thm_subspace_approx_2}. By (\ref{eq_proj_multi_bounded}), we observe that
\e\begin{split}
||\bm{P}_{\bm{\Psi}}\bm{u}_{N,\mathbb{W}}^{(l)}||_2 &\geq ||\overline{\bm{\Psi}}~\overline{\bm{\Psi}}^H\bm{u}_{N,\mathbb{W}}^{(l)}||_2\\
& = ||\sum_{i=0}^{J-1}\sum_{l_i=0}^{k_i-1} \bm{E}_{f_i}\bm{s}_{N,W_i}^{(l_i)}\lambda_{N,W_i}^{(l_i)}(\bm{s}_{N,W_i}^{(l_i)})^H \bm{E}_{f_i}^H\bm{u}_{N,\mathbb{W}}^{(l)}||_2\\
& = ||\bm{B}_{N,\mathbb{W}}\bm{u}_{N,\mathbb{W}}^{(l)}- \sum_{i=0}^{J-1}\sum_{l_i=k_i}^{N-1} \bm{E}_{f_i}\bm{s}_{N,W_i}^{(l_i)}\lambda_{N,W_i}^{(l_i)}(\bm{s}_{N,W_i}^{(l_i)})^H \bm{E}_{f_i}^H\bm{u}_{N,\mathbb{W}}^{(l)}||_2\\
&\geq ||\bm{B}_{N,\mathbb{W}}\bm{u}_{N,\mathbb{W}}^{(l)}||_2 - \sum_{i=0}^{J-1}\sum_{l_i=k_i}^{N-1}||  \bm{E}_{f_i}\bm{s}_{N,W_i}^{(l_i)}\lambda_{N,W_i}^{(l_i)}(\bm{s}_{N,W_i}^{(l_i)})^H \bm{E}_{f_i}^H\bm{u}_{N,\mathbb{W}}^{(l)}||_2\\
&\geq \lambda_{N,\mathbb{W}}^{(l)} - \sum_{i=0}^{J-1}\sum_{l_i=k_i}^{N-1}\lambda_{N,W_i}^{(l_i)}.~~~ \square
\end{split}\nonumber\ee

\section{Proof of Corollary~\ref{cor_subspace_approx_3}} \label{proof_cor_subspace_approx_3}

\noindent{\textbf{Proof.}} It follows from Theorem \ref{thm_subspace_approx_2} that
\begin{equation}\begin{split}
||\bm{P}_{\bm{\Psi}}\bm{u}_{N,\mathbb{W}}^{(l)}||_2 &\geq \lambda_{N,\mathbb{W}}^{(l)} - \sum_{i=0}^{J-1}\sum_{l_i=k_i}^{N-1}\lambda_{N,W_i}^{(l_i)}\\
&\geq 1 - \overline{C}_1(\mathbb{W},\epsilon)N^2e^{-\overline{C}_2(\mathbb{W},\epsilon)N} - \sum_{i=0}^{J-1}\sum_{l_i=k_i}^{N-1}C_3(W_i,\epsilon)e^{-C_4(W_i,\epsilon)N}\\
&\geq 1 - \overline{C}_1(\mathbb{W},\epsilon)N^2e^{-\overline{C}_2(\mathbb{W},\epsilon)N} - \sum_{i=0}^{J-1}\sum_{l_i=k_i}^{N-1}\frac{1}{J}\overline{C}_3(\mathbb{W},\epsilon)e^{-\overline{C}_4(\mathbb{W},\epsilon)N}\\
&\geq 1 - \overline{C}_1(\mathbb{W},\epsilon)N^2e^{-\overline{C}_2(\mathbb{W},\epsilon)N} - N\overline{C}_3(\mathbb{W},\epsilon)e^{-\overline{C}_4(\mathbb{W},\epsilon)N}
\end{split}\nonumber\end{equation}
for all $N\geq \max\{\overline{N}_0(\mathbb{W},\epsilon),\overline{N}_1(\mathbb{W},\epsilon)\}$,
where the second line follows by bounding the $\lambda_{N,\mathbb{W}}^{(l)}$ term using Theorem~\ref{thm:EigConcentrationMulitband}  and by bounding the $\lambda_{N,W_i}^{(l_i)}$ terms using Lemma \ref{lem:EigClusteringDPSS}, and the third line follows because $\overline{C}_3(\mathbb{W},\epsilon)  = J\max{\{C_{3}(W_i,\epsilon), ~\forall~ i \in[J]\}}$ and $\overline{C}_4(\mathbb{W},\epsilon) = \min{\{C_{4}(W_i,\epsilon), ~\forall~ i \in[J]\}}$.

Let $\kappa_2(N,\mathbb W,\epsilon) =\overline{C}_1(\mathbb{W},\epsilon)N^2e^{-\overline{C}_2(\mathbb{W},\epsilon)N}+ N\overline{C}_3(\mathbb{W},\epsilon)e^{-\overline{C}_4(\mathbb{W},\epsilon)N}$. Then $||\bm{u}_{N,\mathbb{W}}^{(l)} - \bm{P}_{\bm{\Psi}}\bm{u}_{N,\mathbb{W}}^{(l)}||_2^2\leq 2\kappa_2(N,\mathbb W,\epsilon) - \kappa_2^2(N,\mathbb W,\epsilon)$. Noting also that $\langle \bm{u}_{N,\mathbb{W}}^{(l)}, \bm{u}_{N,\mathbb{W}}^{(k)}\rangle = 0$ for all $k\neq l$, (\ref{eq:subsapceAngle2-2}) follows directly from Lemma~\ref{lem:BoundSubspaceAngle}. ~~~$\square$

\section{DTFT of DPSS vectors}\label{sec:DTFTofDPSS}

\begin{figure}[t]
\centering
\includegraphics[width=4in]{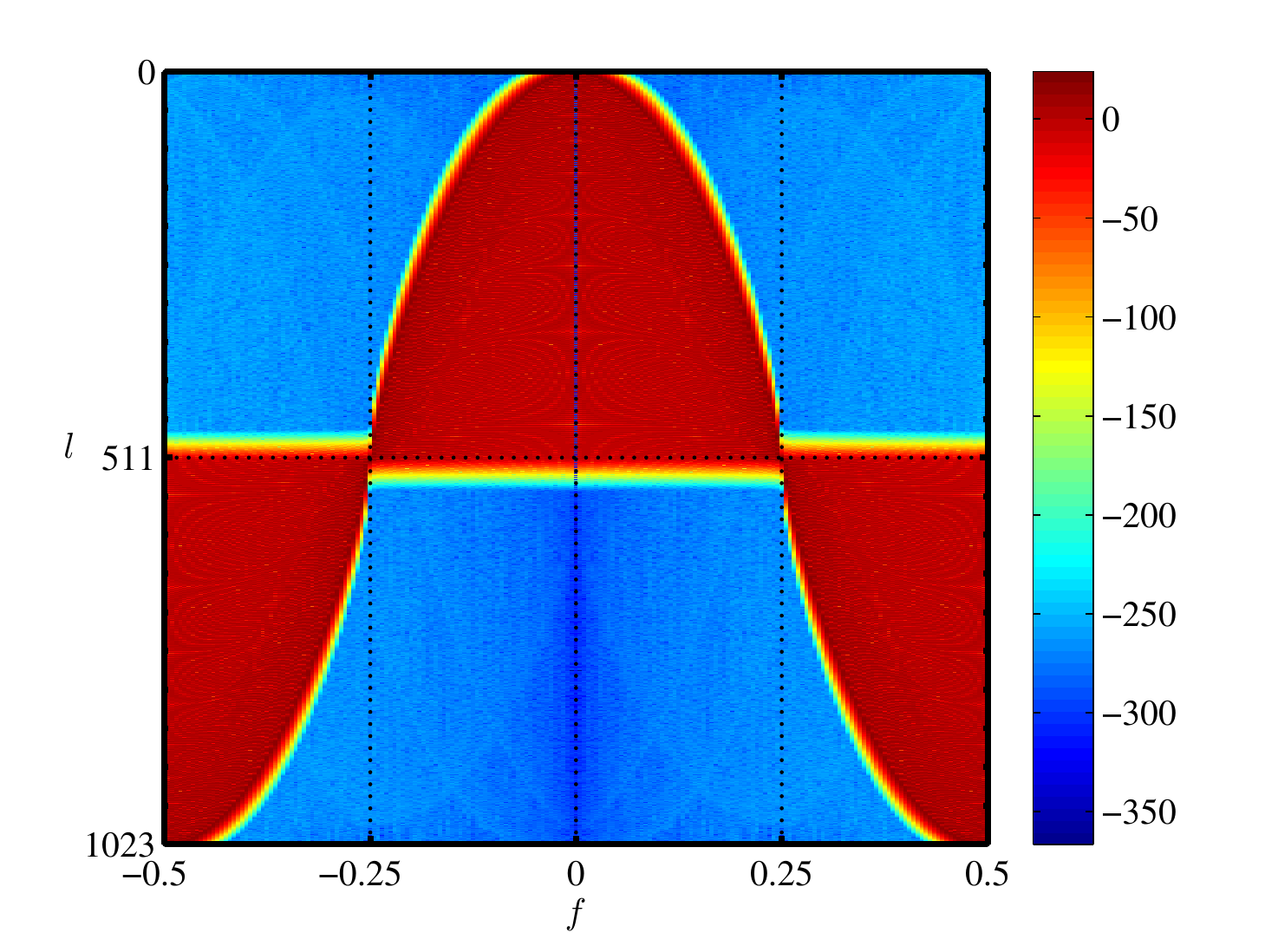}
\vspace{-0.5cm} \caption{\small\sl Illustration of $\left|\widetilde{\bm{s}}_{N,W}^{(l)}(f)\right|^2$, or the energy in $\{\bm{e}_f\}$ captured by each DPSS vector. The horizontal axis stands for the digital frequency $f$, which ranges over $[-\frac{1}{2},\frac{1}{2}]$, while the vertical axis stands for the index $l\in[N]$. The $l$-th horizontal line shows $10\log_{10}\left|\widetilde{\bm{s}}_{N,W}^{(l)}(f)\right|^2$. Here $N = 1024$ and $W= \frac{1}{4}$.
}\label{figure-1}
\end{figure}

The results presented in this appendix are useful in Appendix~\ref{proof:DPSSApproxPureTone}, where we analyze the performance of the DPSS vectors for representing sampled pure tones inside the band of interest. Let $\widetilde{\bm{s}}_{N,W}^{(l)}(f)$ denote the DTFT of the sequence $\mathcal{T}_N(s_{N,W}^{(l)})$, i.e., $\widetilde{\bm{s}}_{N,W}^{(l)}(f) = \sum_{n=0}^{N-1}s_{N,W}^{(l)}[n]e^{-j2\pi fn}$. Figure~\ref{figure-1} shows $\widetilde{\bm{s}}_{N,W}^{(l)}(f)$ for all $l\in[N]$ with $N=1024$ and $W = \frac{1}{4}$. We observe that the first $\approx 2NW$ DPSS vectors have their spectrum mostly concentrated in $[-W,W]$, only a small fraction of DPSS vectors whose indices are near $2NW$ have a relatively flat spectrum over $[-\frac{1}{2},\frac{1}{2}]$, and the remaining DPSS vectors have their spectrum mostly concentrated outside of the band $[-W,W]$. This phenomenon is captured formally in the asymptotic expressions for $\lambda_{N,W}^{(l)}$ and {\color{red} $\widetilde{\bm{s}}^{(l)}_{N,W}(f)$} from~\cite{Slepian78DPSS}.

\begin{lem}(\cite{Slepian78DPSS}) Fix $W\in(0,\frac{1}{2})$ and $\epsilon\in(0,1)$. Let $\alpha := 1-A = 1-\cos 2\pi W$.
\begin{enumerate}
\item For fixed $l$, as $N \rightarrow \infty$, we have
$$
1-\lambda_{N,W}^{(l)}\sim c_5^2/\left(2\sqrt{2\alpha}\right)
$$
and
$$
\widetilde{\bm{s}}_{N,W}^{(l)}(f)\sim\left\{\begin{array}{ll}c_3f_4(f),& W\leq |f|\leq \arccos(A-N^{-3/2})/2\pi,\\
c_5f_5(f), & \arccos(A-N^{-3/2})/2\pi\leq |f|\leq 1/2.
\end{array}\right.
$$
Here \begin{equation}
\begin{split}
c_5 &= (l!)^{-1/2}\pi^{1/4}2^{(14l+15)/8}\alpha^{(2l+3)/8}N^{(2l+1)/4}(\sqrt{2}+\sqrt{\alpha})^{-N}(2-\alpha)^{(N-l-1/2)/2}\\
&= (l!)^{-1/2}\pi^{1/4}2^{(14l+15)/8}\alpha^{(2l+3)/8}N^{(2l+1)/4}(2-\alpha)^{-(l+1/2)/2}e^{-\frac{\gamma}{2} N},\\
c_3 &= \pi^{1/2}2^{-1/2}\alpha^{-1/4}[2-\alpha]^{-1/4}N^{1/2}c_5 = O(N^{1/2}) c_5, \\
  \gamma &= \log(1+\frac{2\sqrt\alpha}{\sqrt2-\sqrt\alpha}),\\
f_4(f) &= J_0\left(\frac{N}{\sqrt{2-\alpha}}\sqrt{A-\cos\left(2\pi f\right)}\right),\\ f_5(f) &=\frac{\cos\left(\frac{N}{2}\arcsin\left(\theta(f)\right)+\frac{1}{2}(l+\frac{1}{2})\arcsin\left(\phi(f)\right)+(l-N)\frac{\pi}{4}+\frac{3\pi}{8}\right)}{\left((A-\cos \left(2\pi f\right))(1-\cos \left(2\pi f\right))\right)^{1/4}},\\
 \theta(f) & = \frac{\alpha+2\cos \left(2\pi f\right)}{2-\alpha}, ~\phi(f) = \frac{(2-3\alpha)-(2+\alpha)\cos \left(2\pi f\right)}{(2-\alpha)(1-\cos \left(2\pi f\right))},
\end{split}\nonumber\end{equation}
where $J_0$ is the Bessel function of the first kind.
\item As $N \rightarrow \infty$ and with $l=\lfloor 2NW(1-\epsilon')\rfloor$ for any $\epsilon'\in(0,\epsilon]$, we have
$$
1-\lambda_{N,W}^{(l)}\sim 2\pi L_2^{-1}d_6^2
$$
and
$$
\widetilde{\bm{s}}_{N,W}^{(l)}(f)\sim\left\{\begin{array}{ll}d_4g_5(f),& W\leq |f|\leq \arccos(A-N^{-1})/2\pi,\\
d_6g_6(f), & \arccos(A-N^{-1})/2\pi\leq |f|\leq 1/2.
\end{array}\right.
$$
Here \begin{equation}
\begin{split}
d_6 &= (L_2)^{-1/2}\pi^{1/2}2^{1/2}e^{-CL_4/4}e^{-NL_3/2}, \\
d_4 &= (L_2)^{-1/2}\pi(1-A^2)^{-1/4}e^{-CL_4/4}e^{-NL_3/2}N^{1/2},\\
g_5(f) &= J_0\left(N\sqrt{\frac{B-A}{1-A^2}\left(\cos(2\pi f)-A\right)}\right),\\
 g_6(f) &= R(f)\cos\left(\pi N\int_{f}^{1/2}\sqrt{\frac{B-\cos(2\pi t)}{A- \cos(2\pi t)}}dt + \frac{\pi C}{2}\int_f^{1/2}\frac{dt}{\sqrt{\left(B - \cos(2\pi t)\right)\left(A - \cos(2\pi t)\right)}} + \theta \right),\\
 R(f) & = \left|\left(B - \cos(2\pi f)\right)\left(A - \cos(2\pi f) \right)\right|^{-1/4}, ~ C=\frac{1}{L_2}\mod\left(\frac{N}{2}L_1 + \left(2+(-1)^l\right)\frac{\pi}{4},2\pi\right),\\
 \theta &= \mod\left(\frac{\pi}{4}-\frac{N}{2}L_5-\frac{C}{4}L_6,2\pi\right),\\
 L_1 &= \int_B^1 P(\xi)d\xi,~L_2 = \int_B^1 Q(\xi)d\xi,~L_3 = \int_A^B P(\xi)d\xi,~L_4 = \int_A^B Q(\xi)d\xi,~L_5 = \int_{-1}^A P(\xi)d\xi,~L_6 = L_2,\\
 P(\xi) &= \left|\frac{\xi - B}{\left(\xi - A\right)\left(1-\xi^2\right)} \right|^{1/2},~Q(\xi) = \left|\left(\xi - B\right)\left(\xi - A\right)\left(1-\xi^2\right)\right|^{-1/2},
 \end{split}\nonumber\end{equation}
where $B$ is determined so that $\int_B^1\sqrt{\frac{\xi - B}{\left(\xi - A\right)\left(1 - \xi^2\right)}}d\xi = \frac{l}{N}\pi$ and $\mod(y,2\pi)$ returns the remainder after division of $y$ by $2\pi$.
\end{enumerate}
\label{lem:DTFTdpssSlepian}
\end{lem}

\section{Proof of Theorem \ref{thm:DPSSApproxPureTone}} \label{proof:DPSSApproxPureTone}

Noting that $\bm{S}_{N,W}$ forms an orthobasis for $\mathbb{C}^{N\times N}$, the main idea is to show that the DPSS vectors $\bm{s}_{N,W}^{(2NW(1+\epsilon))}, \bm{s}_{N,W}^{(2NW(1+\epsilon)+1)}, \ldots,\bm{s}_{N,W}^{(N-1)}$ have their spectrum most concentrated outside of the band $[-W,W]$.

Since the sequence $s_{N,W}^{(l)}$ is exactly bandlimited to the frequency range $|f|\leq W$, we know that its DTFT $\widetilde{s}_{N,W}^{(l)}(f):=\sum_{n=-\infty}^\infty s_{N,W}^{(l)}[n]e^{j2\pi fn}$ vanishes for all $W<|f|<\frac{1}{2}$. By noting that the first $\approx 2NW$ DPSS's are also approximately time-limited to the index range $n=0,1,\ldots,N-1$, we may expect that $\widetilde{\bm{s}}_{N,W}^{(l)}(f):=\sum_{n=0}^{N-1} \bm{s}_{N,W}^{(l)}[n]e^{j2\pi fn}$ is also approximately $0$ for all $W<|f|<\frac{1}{2}$ and $l\leq 2NW(1-\epsilon)$. This illustrates informally why the DTFT of the first $\approx 2NW$ DPSS vectors is concentrated inside the band $[-W,W]$. By employing the antisymmetric property~\cite{Slepian78DPSS} which states that $|\widetilde{\bm{s}}_{N,W}^{(l)}(f)| = |\widetilde{\bm{s}}_{N,\frac{1}{2}-W}^{(N-1-l)}(\frac{1}{2}-f)|$, we then have that the DPSS vectors $\bm{s}_{N,W}^{(2NW(1+\epsilon))}$, $\bm{s}_{N,W}^{(2NW(1+\epsilon)+1)}$, $\ldots,\bm{s}_{N,W}^{(N-1)}$ are almost orthogonal to any  sinusoid with frequency inside the band $[-W,W]$.

Recall that $\widetilde{\bm{s}}_{N,W}^{(l)}(f)$ is the DTFT of the sequence $\mathcal{T}_N(s_{N,W}^{(l)})$, i.e., $\widetilde{\bm{s}}_{N,W}^{(l)}(f) = \sum_{n=0}^{N-1}s_{N,W}^{(l)}[n]e^{-j2\pi fn}$. We have
$$
\langle \bm{s}_{N,W}^{(l)}, \bm{e}_{f}\rangle = \widetilde{\bm{s}}_{N,W}^{(l)}(f),
$$
for all $l\in[N]$. As we have observed in Figure~\ref{figure-1}, the spectrum of the first $\approx 2NW$ DPSS vectors is approximately concentrated on the frequency interval $[-W,W]$.  This behavior is captured formally in the following results.

\begin{cor} Let $A = \cos 2\pi W$. For fixed $W\in(0,\frac{1}{2})$ and $\epsilon\in(0,\min(\frac{1}{2W}-1,1))$, there exists a constant $C_6(W,\epsilon)$ (which may depend on $W$ and $\epsilon$) such that
$$
|\widetilde{\bm{s}}_{N,W}^{(l)}(f)|\leq C_6(W,\epsilon)N^{3/4}e^{-\frac{C_2(W,\epsilon)}{2}N},~~  W\leq |f|\leq 1/2
$$
 for all $N\geq N_0(W,\epsilon)$ and $l\leq 2NW(1-\epsilon)$. Here $C_2(W,\epsilon)$ and $N_0(N,\epsilon)$ are constants specified in Lemma \ref{lem:EigClusteringDPSS}.
\label{lem:DTFT_DPSS_1}\end{cor}
\vspace{0.3cm}
\noindent{\textbf{Proof}} (of Corollary \ref{lem:DTFT_DPSS_1}).

The main approach is to bound $\widetilde{\bm s}_{N,W}^{(l)}(f),~ W\leq|f|\leq 1/2$  with the expressions presented in Lemma~\ref{lem:DTFTdpssSlepian}. Suppose $\epsilon\in(0,1)$ is fixed.

\begin{enumerate}
\item For fixed $l$ and large $N$:

In order to quantify the decay rate of $|\widetilde{\bm{s}}_{N,W}^{(l)}(f)|$, we exploit some results concerning of $f_4(f)$ from~\cite{Olver:2010:NIST} and $f_5(f)$ as follows:
\e|J_0(x)|\leq 1, ~\forall~ x\geq 0,\label{eq:boundBessel}\ee
and for any $\frac{\arccos(A-N^{-3/2})}{2\pi}\leq |f|\leq 1/2$, one may verify that
\begin{equation}
\begin{split}
|f_5(f)|&\leq \frac{1}{\left((A-\cos \left(2\pi f\right))(1-\cos \left(2\pi f\right))\right)^{1/4}}\\
&\leq \frac{1}{\left((A-\left(A-N^{-3/2})\right)(1-\left(A-N^{-3/2})\right)\right)^{1/4}}\\
&\leq \frac{1}{\left((N^{-3/2}))(N^{-3/2}))\right)^{1/4}} = N^{3/4},
\end{split}\nonumber\end{equation}
where the last line follows because $1-A\geq 0$.

Recall that $c_3 = \pi^{1/2}2^{-1/2}\alpha^{-1/4}\left(2-\alpha\right)^{-1/4}N^{1/2}c_5$ and $c_5 \sim \sqrt{2\sqrt{2\alpha}\left(1-\lambda_{N,W}^{(l)}\right)}$.
Plugging these into Lemma \ref{lem:DTFTdpssSlepian} and utilizing Lemma \ref{lem:EigClusteringDPSS}, we get the exponential decay of $|\widetilde{\bm{s}}_{N,W}^{(l)}(f)|$, $|f|\geq W$  as
$$
|\widetilde{\bm{s}}_{N,W}^{(l)}(f)|\leq\left\{\begin{array}{ll}C_7'(W,\epsilon)N^{1/2}e^{-\frac{C_2}{2}N},& W\leq |f|\leq \arccos\left(A-N^{-3/2}\right)/2\pi,\\
C_8'(W,\epsilon)N^{3/4}e^{-\frac{C_2}{2}N}, & \arccos\left(A-N^{-3/2}\right)/2\pi\leq |f|\leq 1/2,
\end{array}\right.$$
for fixed $l$ and $N\geq N_0(W,\epsilon)$. Here $C_7'(W,\epsilon) = \pi^{1/2}2^{1/4}\left(2-\alpha\right)^{-1/4}\sqrt{C_1(W,\epsilon)}$,  $C_8'(W,\epsilon) = (2\sqrt{2\alpha}C_1(W,\epsilon))^{1/2}$, and $N_0(W,\epsilon)$, $C_1(W,\epsilon)$ and $C_2(W,\epsilon)$ are constants as specified in Lemma \ref{lem:EigClusteringDPSS}.

\item For large $N$ and $l=\lfloor 2NW(1-\epsilon')\rfloor, ~\forall~\epsilon'\in(0,\epsilon]$:

Note that $\int_B^1\sqrt{\frac{\xi - B}{\left(\xi - A\right)\left(1 - \xi^2\right)}}d\xi$ is a decreasing function of $B$ and $\int_A^1\sqrt{\frac{\xi - A}{\left(\xi - A\right)\left(1 - \xi^2\right)}}d\xi = 2W\pi>\frac{l}{N}\pi$. Hence $1>B>A$. Now we have
$$
|g_6(f)|\leq |R(f)|\leq \frac{1}{\left(A-\cos(2\pi f)\right)^{1/2}}\leq \frac{1}{\left(A-(A-N^{-1})\right)^{1/2}}\leq N^{1/2}
$$
for all $\arccos(A-N^{-1})/2\pi\leq |f|\leq 1/2$.

Recall that $\left|g_5(f)\right|\leq = 1$ from (\ref{eq:boundBessel}), $d_4 = \pi^{1/2}(1-A^2)^{-1/4}2^{-1/2}N^{1/2}d_6$ and $d_6 \sim \sqrt{\frac{1-\lambda_{N,W}^{(l)}}{2\pi}}$. Plugging these into Lemma \ref{lem:DTFTdpssSlepian} and utilizing the bound on $\lambda_{N,W}^{(l)}$ in Lemma \ref{lem:EigClusteringDPSS}, we get the exponential decay of $|\widetilde{\bm{s}}_{N,W}^{(l)}(f)|$, $|f|\geq W$  as
$$
|\widetilde{\bm{s}}_{N,W}^{(l)}(f)|\leq\left\{\begin{array}{ll}C_7''(W,\epsilon)N^{1/2}e^{-\frac{C_2}{2}N},& W\leq |f|\leq \arccos[A-N^{-1}]/2\pi,\\
C_8''(W,\epsilon)N^{1/2}e^{-\frac{C_2}{2}N}, & \arccos[A-N^{-1}]/2\pi\leq |f|\leq 1/2,
\end{array}\right.$$
for all $l=\lfloor 2NW(1-\epsilon')\rfloor, ~\forall~\epsilon'\in(0,\epsilon]$ and $N\geq N_0(W,\epsilon)$. Here $C_8''(W,\epsilon) = \sqrt{C_1(W,\epsilon)/2\pi}$, $C_7''(W,\epsilon) = 2^{-1}(1-A^2)^{-1/4}\sqrt{C_1(W,\epsilon)}$,  and $N_0(W,\epsilon)$, $C_1(W,\epsilon)$ and $C_2(W,\epsilon)$ are constants as specified in Lemma \ref{lem:EigClusteringDPSS}.
\end{enumerate}
Set $$C_6(W,\epsilon) = \max\left\{C_7'(W,\epsilon),C_8'(W,\epsilon),C_7''(W,\epsilon),C_8''(W,\epsilon)\right\} = \max\left\{\pi^{1/2}\left(\frac{2}{2-\alpha}\right)^{1/4},2^{-1}(1-A^2)^{-1/4}\right\}\sqrt{C_1(W,\epsilon)}.$$
This completes the proof of Corollary \ref{lem:DTFT_DPSS_1}.~~$\square$

\begin{lem}(\cite{Slepian78DPSS})
For fixed $W\in(0,\frac{1}{2})$ and $\epsilon\in(0,\frac{1}{2W}-1)$, $\widetilde{\bm{s}}_{N,W}^{(l)}(f)$ and $\widetilde{\bm{s}}_{N,\frac{1}{2}-W}^{(N-1-l)}(f)$ satisfy
$$
|\widetilde{\bm{s}}_{N,W}^{(l)}(f)| = |\widetilde{\bm{s}}_{N,\frac{1}{2}-W}^{(N-1-l)}(\frac{1}{2}-f)|
$$
for all $l\geq 2NW(1+\epsilon)$.
\label{lem_dpss_spectrum_symmetry}\end{lem}
Now we can conclude that $\langle \bm{e}_{f}, \bm{s}_{N,W}^{(l)} \rangle $ decays exponentially in $N$ for all $l\geq 2NW(1+\epsilon)$ and $|f|\leq W$ by combining the above results.

\begin{cor}
Fix $W\in(0,\frac{1}{2})$ and $\epsilon\in(0,\frac{1}{2W}-1)$. Let $W' = \frac{1}{2} - W$ and $\epsilon' = \frac{W}{\frac{1}{2}-W}\epsilon$. Then
$$
|\langle \bm{e}_{f}, \bm{s}_{N,W}^{(l)} \rangle| = |\widetilde{\bm{s}}_{N,W}^{(l)}(f)|\leq C_6(W',\epsilon')N^{3/4}e^{-\frac{C_2(W',\epsilon')}{2}N},~\forall |f|\leq W~
$$
for all $N\geq N_0(W',\epsilon')$ and all $l\geq 2NW(1+\epsilon)$. Here, $C_2(W',\epsilon')$ and $N_0(W',\epsilon')$ are constants specified in Lemma \ref{lem:EigClusteringDPSS} with respect to $W'$ and $\epsilon'$, and $C_6(W',\epsilon')$ is the constant specified in Corollary~\ref{lem:DTFT_DPSS_1} with respect to $W'$ and $\epsilon'$.
\label{cor:DTFT_DPSS}\end{cor}
\vspace{0.3cm}
\noindent{\textbf{Proof of Corollary \ref{cor:DTFT_DPSS}.}} Let $l' = N - 1 - l$. For all $l\geq 2NW(1+\epsilon)$, we have
$$l' = N-1-l \leq N-2NW(1+\epsilon) = 2N(\frac{1}{2}-W)(1-\frac{W}{\frac{1}{2}-W}\epsilon).$$
Let $W' = \frac{1}{2} - W$ and $\epsilon' = \frac{W}{\frac{1}{2}-W}\epsilon\in (0,1)$. It follows from from Corollary \ref{lem:DTFT_DPSS_1} and Lemma \ref{lem_dpss_spectrum_symmetry} that $$|\langle \bm{e}_{f}, \bm{s}_{N,W}^{(l)} \rangle| = |\langle \bm{e}_{\frac{1}{2}-f}, \bm{s}_{N,W'}^{(l')} \rangle|\leq C_6(W',\epsilon')N^{3/4}e^{-\frac{C_2(W',\epsilon')}{2}N},~\forall~|f|\leq W$$
for all $N\geq N_0(W',\epsilon')$. ~~~$\square$

\vspace{0.3cm}
Recall that
$C_6(W',\epsilon') = \max\left\{\pi^{1/2}\left(\frac{2}{\alpha}\right)^{1/4},2^{-1}(1-A^2)^{-1/4}\right\}\sqrt{C_1(W',\epsilon')}$ with $A = \cos(2\pi W)$ and $\alpha = 1-A$. As $W$ gets closer to $0$ or $\frac{1}{2}$, the variable $(1-A^2)^{-1/4}$ becomes larger, and we have $(1-A^2)^{-1/4} \rightarrow 1/\sqrt{2\pi W}$ as $W \rightarrow0$. Also we have $\left(\frac{2}{\alpha}\right)^{1/4} \rightarrow 1/\sqrt{\pi W}$ as $W \rightarrow0$.
Therefore, for any non-negligible bandwidth which is the main assumption in this paper, the variable $\max\left\{\pi^{1/2}\left(\frac{2}{\alpha}\right)^{1/4},2^{-1}(1-A^2)^{-1/4}\right\}\sqrt{C_1(W',\epsilon')}$ would not be too large.

Now, for fixed $W\in(0,\frac{1}{2})$ and $\epsilon\in(0,\frac{1}{2W}-1)$, we have
\e\begin{split}
||\bm{e}_f - \bm{P}_{[\bm{S}_{N,W}]_k}\bm{e}_f ||_2^2 &= \sum_{l=2NW(1+\epsilon)}^{N-1} |\langle \bm{e}_{f}, \bm{s}_{N,W}^{(l)} \rangle|^2 \\
& \leq \sum_{l=2NW(1+\epsilon)}^{N-1} C_6^2(W',\epsilon')N^{3/2}e^{-C_2(W',\epsilon')N}\\
& \leq C_9(W',\epsilon')N^{5/2}e^{-C_2(W',\epsilon')N}
\end{split}\nonumber\ee
for all $|f|\leq W$ and $N\geq N_0(W',\epsilon')$, where $C_9(W',\epsilon') = C_6^2(W',\epsilon')$. ~~~$\square$

\section{Proof of Corollary~\ref{cor:MDPSSApproxPureTone}} \label{proof:corMDPSSApproxPureTone}

\noindent{\textbf{Proof.}} Suppose $f\in[f_i - W_i, f_i + W_i]$ for some particular $i\in[J]$. Let $C_{10}(\mathbb{W},\epsilon) = \max\{C_{9}(W_i',\epsilon'),\forall i\in[J]\}$ and $C_{11}(\mathbb{W},\epsilon) = \min\{C_{2}(W_i',\epsilon'),\forall i\in[J]\}$. It follows from Theorem~\ref{thm:DPSSApproxPureTone} that
\e\begin{split}
||\bm{e}_f - \bm{P}_{\bm{\Psi}}\bm{e}_f||_2^2 &\leq ||\bm{e}_f - \bm{P}_{[\bm{E}_{f_i}\bm{S}_{N,W_i}]_{2NW_i(1+\epsilon)}}\bm{e}_f||_2^2\\
&=||\bm{e}_{f-f_i} - \bm{P}_{[\bm{S}_{N,W_i}]_{2NW_i(1+\epsilon)}}\bm{e}_{f-f_i}||_2^2\\
& \leq C_9(W_i',\epsilon')N^{5/2}e^{-C_2(W_i',\epsilon')N}\leq C_{10}(\mathbb{W},\epsilon)N^{5/2}e^{-C_{11}(\mathbb{W},\epsilon)N}
\end{split}\nonumber\ee
for all $N\geq N_0(W_i',\epsilon')$. We complete the proof by setting $N_2(\mathbb{W},\epsilon) = \max\{N_0(W_i',\epsilon'),\forall i\in[J]\}$. ~~~$\square$

\section{ Proof of Theorem \ref{thm_appr_uniform}}\label{proof_thm_appr_uniform}

\noindent\textbf{Proof.}
Since $\bm{x}_0, \bm{x}_1, \ldots, \bm{x}_{J-1}$ are independent and zero-mean, we have
$$\mathbb{E}\left[\left\|\bm{x}\right\|_2^2\right]=\sum_{n=0}^{N-1}\mathbb{E}\left[\left|\bm{x}[n]\right|^2\right]=\sum_{n=0}^{N-1}\sum_{0\leq i,i'\leq J-1}\mathbb{E}\left[\bm{x}_i[n]\overline{\bm{x}_i'[n]}\right]=\sum_{n=0}^{N-1}\sum_{i=0}^{J-1}\mathbb{E}\left[\left|\bm{x}_i[n]\right|^2\right]
=N\sum_{i=0}^{J-1} \frac{1}{J}=N.$$
Applying Theorem~\ref{thm:DPSSApproxBandpass}, we acquire
$$\mathbb{E}\left[\left\|\bm{x}_i-\bm{P}_{\left[\bm{E}_{f_i}\bm{S}_{N,W_i}\right]_{k_i}}\bm{x}\right\|_2^2\right]= \frac{1}{|\mathbb{W}|}\sum_{l=k_i}^{N-1}\lambda_{N,W_i}^{(l)}.$$
Note that the power spectrum $P_{x_i}(F)$ assumed in (\ref{eq:powerspectrumMD}) results in the constant $\frac{1}{|\mathbb{W}|}$ instead of $\frac{1}{2W_i}$.

Now, we have
\begin{equation}
\begin{split}
\mathbb{E}\left[\left\|\bm{x}-\bm{P}_{\bm{\Psi}}\bm{x}\right\|_2^2\right]&=\mathbb{E}\left[\left\|\sum_{i=0}^{J-1}\bm{x}_i-\bm{P}_{\bm\Psi}(\sum_{i=0}^{J-1}\bm{x}_i)\right\|_2^2\right]=\mathbb{E}\left[\left\|\sum_{i=0}^{J-1}\left(\bm{x}_i-\bm{P}_{\bm\Psi}\bm{x}_i\right)\right\|_2^2\right]\\
&=\mathbb{E}\left[\left(\sum_{i=0}^{J-1}\left(\bm{x}_i-\bm{P}_{\bm\Psi}\bm{x}_i\right)^H\right)\left(\sum_{i=0}^{J-1}\left(\bm{x}_i-\bm{P}_{\bm\Psi}\bm{x}_i\right)\right)\right]\\
&=\mathbb{E}\left[\sum_{i=0}^{J-1}\left\|\bm{x}_i-\bm{P}_{\bm\Psi}\bm{x}_i\right\|_2^2+\sum_{i=0}^{J-1}\sum_{i'=0,i'\neq i}^{J-1}\left(\bm{x}_i-\bm{P}_{\bm\Psi}\bm{x}_i\right)^H\left(\bm{x}_{i'}-\bm{P}_{\bm\Psi}\bm{x}_{i'}\right)\right]\\
&= \sum_{i=0}^{J-1} \mathbb{E}\left[\left\|\bm{x}_i-\bm{P}_{\bm{\Psi}}\bm{x}_i\right\|_2^2\right] + \sum_{i=0}^{J-1}\sum_{i'=0,i'\neq i}^{J-1}\mathbb{E}
\left[\left(\bm{x}_i-\bm{P}_{\bm\Psi}\bm{x}_i\right)^H\left(\bm{x}_{i'}-\bm{P}_{\bm\Psi}\bm{x}_{i'}\right)\right]\\
&= \sum_{i=0}^{J-1} \mathbb{E}\left[\left\|\bm{x}_i-\bm{P}_{\bm\Psi}\bm{x}_i\right\|_2^2\right] + \sum_{i=0}^{J-1}\sum_{i'=0,i'\neq i}^J\mathbb{E}\left[\bm{x}_i^H\bm{x}_{i'}-\bm{x}_i^H\bm{P}_{\bm\Psi}\bm{x}_{i'}\right]\\
&=\sum_{i=0}^{J-1} \mathbb{E}\left[\left\|\bm{x}_i-\bm{P}_{\bm\Psi}\bm{x}_i\right\|_2^2\right]\leq \sum_{i=0}^{J-1} \mathbb{E}\left[\left\|\bm{x}_i-\bm{P}_{[\bm{E}_{f_i}\bm{S}_{N,W_i}]_{k_i}}\bm{x}_i\right\|_2^2\right]\\
& = \sum_{i=0}^{J-1}\frac{1}{|\mathbb{W}|}\sum_{l=k_i}^{N-1}\lambda_{N,W_i}^{(l)}
\end{split}
\nonumber\end{equation}
where the equality in the sixth line follows because $\mathbb{E}\left[\bm{x}_{i'}^H\bm{x}_i\right]=\left(\mathbb{E}\left[\bm{x}_{i'}\right]\right)^H\left(\mathbb{E}\left[\bm{x}_i\right]\right)=0$ and $\mathbb{E}\left[\bm{x}_{i'}^H\bm{P}_{\bm{\Psi}}\bm{x}_i\right]=\left(\mathbb{E}\left[\bm{x}_{i'}\right]\right)^H\left(\mathbb{E}\left[\bm{P}_{\bm{\Psi}}\bm{x}_i\right]\right)=0$ for all $i',i \in [J], i'\neq i$, and the inequality in the sixth line follows because the column space of $[\bm{E}_{f_i}\bm{S}_{N,W_i}]_{k_i}$ is inside the column space of $\bm\Psi$
for all $i\in[J]$.~~ $\square$

\section{Proof of Corollary~\ref{cor:MDPSSApproxMultiband}} \label{proof:corMDPSSApproxMultiband}

\noindent{\textbf{Proof.}}
It is useful to express the sampled bandpass signal $\bm{x}$ as \e\bm{x} = \int_{\mathbb{W}}\widetilde{x}(f)\bm{e}_fdf,\label{eq:SampledMultibandSignal}\ee
where we recall that $\widetilde{x}(f)$ denotes the DTFT of $x[n]$, which is the infinite-length sequence that one obtains by uniformly sampling $x(t)$ with sampling rate $T_s$.

Now it follows from (\ref{eq:SampledMultibandSignal}) that
\e\begin{split}\left\|\bm{x}-\bm{P}_{\bm{\Psi}}\bm{x}\right\|_2^2 &= \left\|\int_{\mathbb{W}}\widetilde{x}(f)\bm{e}_fdf - \int_{\mathbb{W}}\widetilde{x}(f)\bm{P}_{\bm{\Psi}}\bm{e}_fdf\right\|_2^2\\
&=\left\|\int_{\mathbb{W}}\widetilde{x}(f)(\bm{e}_{f} - \bm{P}_{\bm{\Psi}}\bm{e}_{f})df\right\|_2^2\\
&\leq\int_{\mathbb{W}}|\widetilde{x}(f)|^2df \cdot \int_{\mathbb{W}}\|\bm{e}_{f} - \bm{P}_{\bm{\Psi}}\bm{e}_{f}\|_2^2df\\
& \leq \int_{\mathbb{W}}|\widetilde{x}(f)|^2df \cdot  C_{10}(\mathbb{W},\epsilon)N^{5/2}e^{-{C}_{11}(\mathbb{W},\epsilon)N},
\end{split}\nonumber\ee
where the third line follows from the Cauchy-Schwarz inequality and the last line follows from (\ref{eq:MDPSSApproxPureTone}) and the fact that $\int_{\mathbb{W}}\|\bm{e}_{f} - \bm{P}_{\bm{\Psi}}\bm{e}_{f}\|_2^2df\leq |\mathbb{W}|\sup_{f\in\mathbb{W}} \|\bm{e}_{f} - \bm{P}_{\bm{\Psi}}\bm{e}_{f}\|_2^2\leq \sup_{f\in\mathbb{W}} \|\bm{e}_{f} - \bm{P}_{\bm{\Psi}}\bm{e}_{f}\|_2^2$.
~~~$\square$

\bibliographystyle{abbrv}
\bibliography{mybibfileMDPSS}

\end{document}